\documentclass[twocolumn]{aastex6}
\usepackage{natbib}
\usepackage{amsmath,amsthm,amssymb,amsfonts}
\extrafloats{200}

\shorttitle{Transit Signatures of Inhomogeneous Clouds on Hot Jupiters}

\begin{document}

\title{Transit Signatures of Inhomogeneous Clouds on Hot Jupiters: Insights From Microphysical Cloud Modeling}
\author{Diana Powell\altaffilmark{1}, Tom Louden \altaffilmark{2,3}, Laura Kreidberg \altaffilmark{4,5}, Xi Zhang\altaffilmark{6}, Peter Gao\altaffilmark{7,8}, Vivien Parmentier \altaffilmark{9}}

\altaffiltext{1}{Department of Astronomy and Astrophysics, University of California, Santa Cruz, CA 95064; \href{mailto:dkpowell@ucsc.edu}{dkpowell@ucsc.edu}}
\altaffiltext{2}{Department of Physics, University of Warwick, Coventry CV4 7AL, UK}
\altaffiltext{3}{Winton Fellow}
\altaffiltext{4}{Harvard-Smithsonian Center for Astrophysics, Cambridge, MA 02138, USA}
\altaffiltext{5}{Harvard Society of Fellows, Harvard University, Cambridge, MA 02138}
\altaffiltext{6}{Department of Earth and Planetary Sciences, University of California, Santa Cruz, CA 95064}
\altaffiltext{7}{Department of Astronomy, University of California, Berkeley, CA 94720}
\altaffiltext{8}{51 Pegasi b Fellow}
\altaffiltext{9}{Atmospheric, Ocean, and Planetary Physics, Clarendon Laboratory, Department of Physics, University of Oxford, Oxford, OX1 3PU, UK}

\begin{abstract}
 We determine the observability in transmission of inhomogeneous cloud cover on the limbs of hot Jupiters through post processing a general circulation model to include cloud distributions computed using a cloud microphysics model. We find that both the east and west limb often form clouds, but that the different properties of these clouds enhances the limb to limb diffesrences compared to the clear case. Using \textit{JWST} it should be possible to detect the presence of cloud inhomogeneities by comparing the shape of the transit lightcurve at multiple wavelengths because inhomogeneous clouds impart a characteristic, wavelength dependent signature. This method is statistically robust even with limited wavelength coverage, uncertainty on limb darkening coefficients, and imprecise transit times. We predict that the short wavelength slope varies strongly with temperature. The hot limb of the hottest planets form higher altitude clouds composed of smaller particles leading to a strong rayleigh slope. The near infrared spectral features of clouds are almost always detectable, even when no spectral slope is visible in the optical. In some of our models a spectral window between 5 and 9 microns can be used to probe through the clouds and detect chemical spectral features. Our cloud particle size distributions are not log-normal and differ from species to species. Using the area or mass weighted particle size significantly alters the relative strength of the cloud spectral features compared to using the predicted size distribution. Finally, the cloud content of a given planet is sensitive to a species' desorption energy and contact angle, two parameters that could be constrained experimentally in the future.
\end{abstract}

\keywords{planets and satellites: atmospheres -- planets and satellites: gaseous planets} 

\section{Introduction}
Clouds are ubiquitous in the atmospheres of solar system planets and are seemingly abundant in the atmospheres of exoplanets as well, where they affect the atmospheric dynamics, radiative energy distribution, and chemistry. The presence of clouds on exoplanets is commonly inferred through damped spectral features and enhanced Rayleigh-like slopes in the optical \citep[e.g.,][]{2013A&A...559A..33C,2013ApJ...765..127F,2014ApJ...794..155K,2014Natur.505...66K,2014Natur.505...69K,2016ApJ...823..109I,2016Natur.529...59S,2017MNRAS.470..742L} and these effects on the atmospheric spectra strongly inhibit our ability to constrain fundamental atmospheric properties for the majority of exoplanets \citep[e.g.,][]{ackerman-marley-2001,2013ApJ...775...33M,2014ApJ...794..155K,2014Natur.505...69K,2016Natur.529...59S,2018ApJ...860...18P,2018ApJ...863..165G}. An understanding of clouds on exoplanets and their effect on the observed atmospheric spectra is thus essential in interpreting observations.

Transmission spectroscopy is the leading technique for characterization of exoplanet atmospheres \citep[e.g.,][]{seager-sasselov-2000,brown-etal-2001,hubbard-etal-2001}, but its reliance on a slant light path makes it especially sensitive to high altitude clouds \citep{fortney-2005}. Most analysis of transmission spectra use 1D atmospheric models that assume temperature structures, chemical abundances, and cloud particle size distributions that are longitudinally and latitudinally homogeneous \citep[e.g.,][]{2014Natur.505...69K,2015ApJ...815..110M,2016Natur.529...59S}. However, exoplanets are inherently 3D with spectra that may be different at different locations due to differences in temperature structure \citep[e.g.,][]{2016ApJ...829...52F,2019A&A...623A.161C}, atmospheric mixing \citep[e.g.,][]{2012ApJ...751...87D}, cloud properties \citep[e.g.,][]{2015A&A...580A..12L,2016A&A...594A..48L,2018ApJ...860...18P,2019MNRAS.488.1332L}, or a combination of the aforementioned - leading to a globally averaged spectra that is a combination of different spectra from different planetary locations.

\subsection{Inhomogenous Cloud Cover on Hot Jupiters}

Hot Jupiters have particularly inhomogeneous atmospheres because they are highly irradiated by their host stars and are likely tidally-locked which causes them to have strong day-night temperature contrasts. The efficiency of heat redistribution in these atmospheres decreases for planets with higher equilibrium temperatures such that the day-night temperature contrast is particularly extreme for the hottest planets \citep{perez-becker-showman-2013,2016ApJ...821...16K}.

Because cloud properties are highly sensitive to the local atmospheric thermal structure, we expect that these large temperature contrasts will lead to clouds with substantially different masses, vertical distributions, particle size distributions, and compositions \citep{2018ApJ...860...18P}. In particular, there are two identified mechanisms that could give rise to inhomogeneous cloud cover in the atmospheres of hot Jupiters \citep{2016ApJ...820...78L}. The first relies on the meridional transport of cloud particles from the equator to the poles \citep{2013A&A...558A..91P,2015ApJ...813...15C}, leading to equatorial regions with less clouds and a cloud enhancement at the poles. The second relies on significant temperature gradients across the planet that alter the local cloud formation processes, leading to inhomogeneous cloud cover \citep[e.g.,][]{parmentier2016transitions}. 

For many hot Jupiters, the temperature structure on the east limb is substantially hotter than the west limb such that the gaseous species that can condense and form clouds differ significantly \citep{2018ApJ...860...18P}. In particular, hot Jupiters with T$_\text{eq}$ in the range of 1800 -- 2100 K may represent the most dramatic cases of inhomogeneous limb cloud cover. Cooler hot Jupiters may very well exhibit similar inhomogeneity as has been inferred from observations of HD 209458b (T$_\text{eq} \approx 1400$ K) \citep{2017MNRAS.469.1979M}. However, previous work has proposed that atmospheric dynamics may reduce latitudinal and longitudinal inhomogeneities in the cloud properties of cooler hot Jupiters \citep{2017A&A...601A..22L,2018A&A...615A..97L}, complicating the general picture of cloud inhomogeneity.

To date, there are roughly 25 hot Jupiters that have been observed in either transmission, emission, or reflection, within a range of equilibrium temperatures that may form significantly inhomogenous clouds \citep[][etc.]{2018AJ....156..122M,2017ApJ...847L..22F,2018AJ....155..156T,2016Natur.529...59S,2019AJ....157...68B,2018AJ....155...83Z,2013ApJ...776L..25D,2015ApJ...802...51H,2015ApJ...804...94W,2015AJ....150..112S,2014ApJ...785..148R,2015ApJ...811..122W,2017A&A...608A..26M} and nearly all of these planets have spectral signatures that are interpreted as being due to the presence of clouds or hazes. 

\subsection{Finding a Transmission Signature of Inhomogenous Cloud Cover}
Currently, the most robust measure of cloud inhomogeneity is optical phase curves which have offsets (the maximum of the phase curve compared to the secondary eclipse) that can probe longitudinal cloud cover \citep[e.g.,][]{parmentier2016transitions}. Using this method, signatures of inhomogenous cloud cover have been observed in the atmospheres of three hot Jupiters - Kepler-7b, Kepler-12b, and Kepler-41b - the only currently identified planets with optical phase curves that are dominated by atmospheric processes and can be modeled independently of approximations needed to simultaneously model orbital effects \citep{2013ApJ...776L..25D,2015ApJ...802...51H,2015ApJ...804...94W,2015AJ....150..112S,2015PNAS..11213461G}. The presence of inhomogeneous clouds is thus likely common because all of the planets with robust two dimensional atmospheric information have signatures of inhomogeneous clouds \citep{2015AJ....150..112S}. In observations of optical phase curves, however, there can be substantial non-atmospheric processes, such as doppler boosting, tidal ellipsoidal distortion, and planetary obliquity, that require approximations for this form of analysis \citep[see review by][]{2017PASP..129g2001S}. This, coupled with the comparative difficulty of phase curve observations \citep{2017PASP..129g2001S,2018haex.bookE.116P} makes this method of probing inhomogeneous cloud cover difficult to generalize to the vast majority of hot Jupiters. It is therefore of great use to determine a robust observational signature of cloud inhomogenity in transmission alone which, in addition to aiding in planetary characterization, can also be used to constrain models of planetary phase curves. 

Simplified atmospheric modeling has shown that inhomogeneous clouds on the east and west limbs can mimic an atmosphere with high mean molecular weight when observed in transmission \citep{2016ApJ...820...78L}. In addition, single-hemisphere clouds produce significant residuals in the shape of the transit light curve when fitted with a model assuming uniform limb radii \citep{2016ApJ...820...78L,2016A&A...589A..52V}. It has also been suggested that inhomogeneous aerosol coverage could be a diagnostic for distinguishing between clouds and haze in hot Jupiters with T$_\text{eq} \gtrsim 2000$ K \citep{2017ApJ...845L..20K}. However, specific transmission signatures of inhomogeneous cloud cover have not been well constrained. 

In this work we present transmission signatures of inhomogeneous cloud cover that should be observable using the \textit{James Webb Space Telescope} (\textit{JWST}). In Section \ref{cloud_model} we describe our non-equilibrium cloud model in which we determine cloud properties from first principles and discuss our model planet parameters and choice of model grid. In Section \ref{clouds}, we present our derived cloud properties at the relevant locations in the planetary atmosphere for our grid of model hot Jupiters. We calculate transmission spectra using the cloud properties derived from our microphysical model for different planetary locations in our grid in Section \ref{trans} and discuss specific transmission signatures of condensible clouds and their effect on the transmission spectra as a whole. We then present forward and inverse modeling of the light curves of these modeled planets in Section \ref{observability} and present statistically robust transmission metrics of inhomogeneous clouds using \textit{JWST}. We discuss our results in Section \ref{discuss} and present our conclusions in Section \ref{conclude}. 

\section{Cloud Model}\label{cloud_model}

Clouds form via complex microphysical processes that depend strongly on planetary properties, notably a planet's thermal structure, chemical composition, and the strength of mixing in the atmosphere \citep[e.g.,][]{2015A&A...580A..12L,2016A&A...594A..48L,2018ApJ...860...18P,2018ApJ...863..165G}. To model condensible clouds in the atmospheres of hot Jupiters we use the non-equilibrium one dimensional Community Aerosol and Radiation Model for Atmospheres (\textsc{CARMA}) \citep{turco1979,1988JAtS...45.2123T} version 3.0 \citep{JGRD:JGRD15781,JGRD:JGRD14488}. \textsc{CARMA} is a well tested code that was developed to study clouds on Earth and has since been applied to reproduce and understand observations of clouds on Mars \citep{1993Icar..102..261M,1999JGR...104.9043C}, Venus \citep{gao2014bimodal}, Titan \citep{2003Icar..162...94B,2004GeoRL..3117S07B,2006Icar..182..230B}, and Pluto \citep{2017Icar..287..116G}. For a comprehensive discussion of the microphysics and history of \textsc{CARMA} see \citet{2018ApJ...855...86G,turco1979,1988JAtS...45.2123T,JACOBSON1994}. We adapted CARMA to simulate titanium and silicate clouds on hot Jupiters in our previous work \citep{2018ApJ...860...18P}, and in this paper we use an updated version of the model that includes additional cloud species \citep[][also see Appendix \ref{species}]{Gao2019}. We present a brief description of our model setup and refer readers to \citet{2018ApJ...860...18P} and \citet{Gao2019} for a more detailed discussion. 

\begin{deluxetable}{ll}
\tablecolumns{4}
\tablecaption{Model Parameters \label{mps}}
\tablehead{   % column headings
  \colhead{} &
  \colhead{Values} 
}
\startdata
Surface Gravity     & 1000  cm s$^{-2}$ \\  
Atmospheric Mole. Wt. & 2.2 g mol$^{-1}$ (H/He)   \\
T-P Profiles     &  Figure \ref{pt_profs} \\
Vertical Mixing  & Section \ref{kzz_section} \\
Time Step    &  100 s   \\  
Total Simulation Time    &  $10^{9}$ s   \\
Mass Ratio Between Bins  & 2      \\
Number of Bins    & 80      \\
Smallest Bin Size   & 1 nm      \\
\textbf{Boundary Conditions}  &      \\
Clouds (Top)  & Zero Flux      \\
Condensation Nuclei (Top)  & Zero Flux     \\
Condensible gases (Top)  & Zero Flux      \\
Condensible gases (Bottom)  & Solar abundance      \\
Clouds (Bottom)  & 0 cm$^{-3}$     \\
Condensation Nuclei (Bottom)  & 0 cm$^{-3}$     \\
\enddata 
\end{deluxetable} 

\textsc{CARMA} treats the microphysical processes of homogenous nucleation, heterogenous nucleation, condensational growth, evaporation, and coagulation as well as vertical transport of cloud particles due to atmospheric mixing and gravitational settling. For a comprehensive discussion of these processes and the role they play in atmospheres of hot Jupiters see \citet{2018ApJ...860...18P}. \textsc{CARMA} resolves the cloud particle size distribution using bin-scheme microphysics. In the bin-scheme approach, the size distribution is discretized into multiple bins according to size and the particles in each bin evolve freely and interact with other bins in an Eulerian framework. There is no a priori assumption of the particle size distribution. Bin-scheme microphysics is widely used in cloud formation models of Earth's atmosphere and is able to reproduce the multimodal and broad distributions of cloud particles \citep[e.g.,][]{doi:10.1029/2006JD007688,acp-19-1413-2019}. Furthermore, \textsc{CARMA} is a non-equilibrium cloud model such that it simulates the time-dependent formation and evolution of cloud particles. This model can therefore capture subtleties of cloud variability due to microphysical processes. Due to the inherent difference in magnitude between the timescale of atmospheric mixing and the timescales of microphysical processes \citep{2003Icar..162...94B,2018ApJ...860...18P}, our model does indeed predict cloud variability that may be real. However, in this work we present results that are time averaged over this steady state microphysical variability.  

 \begin{figure}[tbp]
\epsscale{1.1}
\plotone{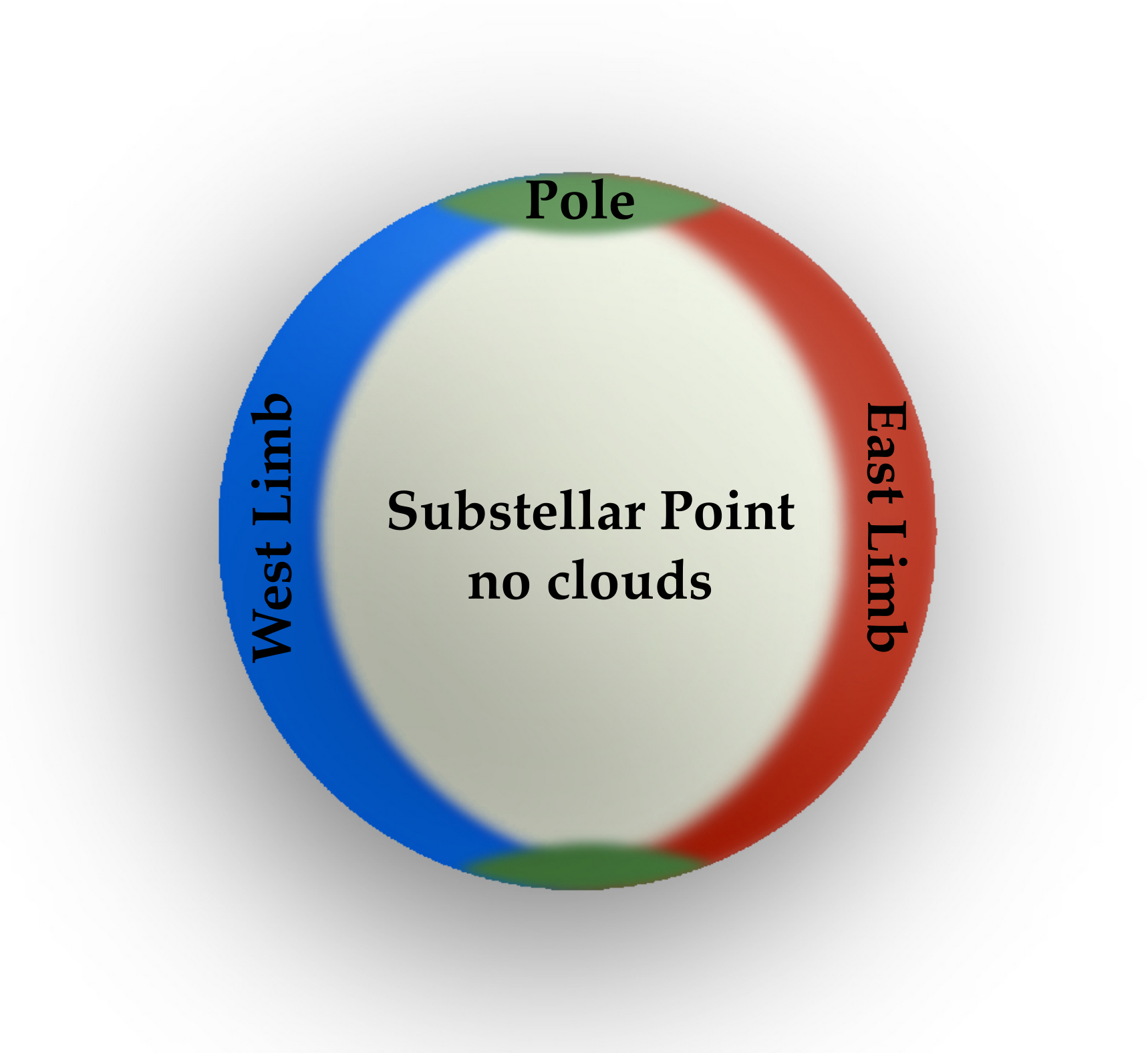}
\caption{A schematic of the atmospheric regions along the terminator of a hot Jupiter that we sample in our modeling: the poles (green), east limb (red), and west limb (blue). For the temperature ranges probed in our modeling we do not expect cloud formation on the dayside \citep{2018ApJ...860...18P}, such that the clouds from the west limb cannot be transported to the east limb along a superrotating equatorial jet.}
\label{model_setup}
\end{figure}

\begin{figure*}[tbp]
\epsscale{1.15}
\plotone{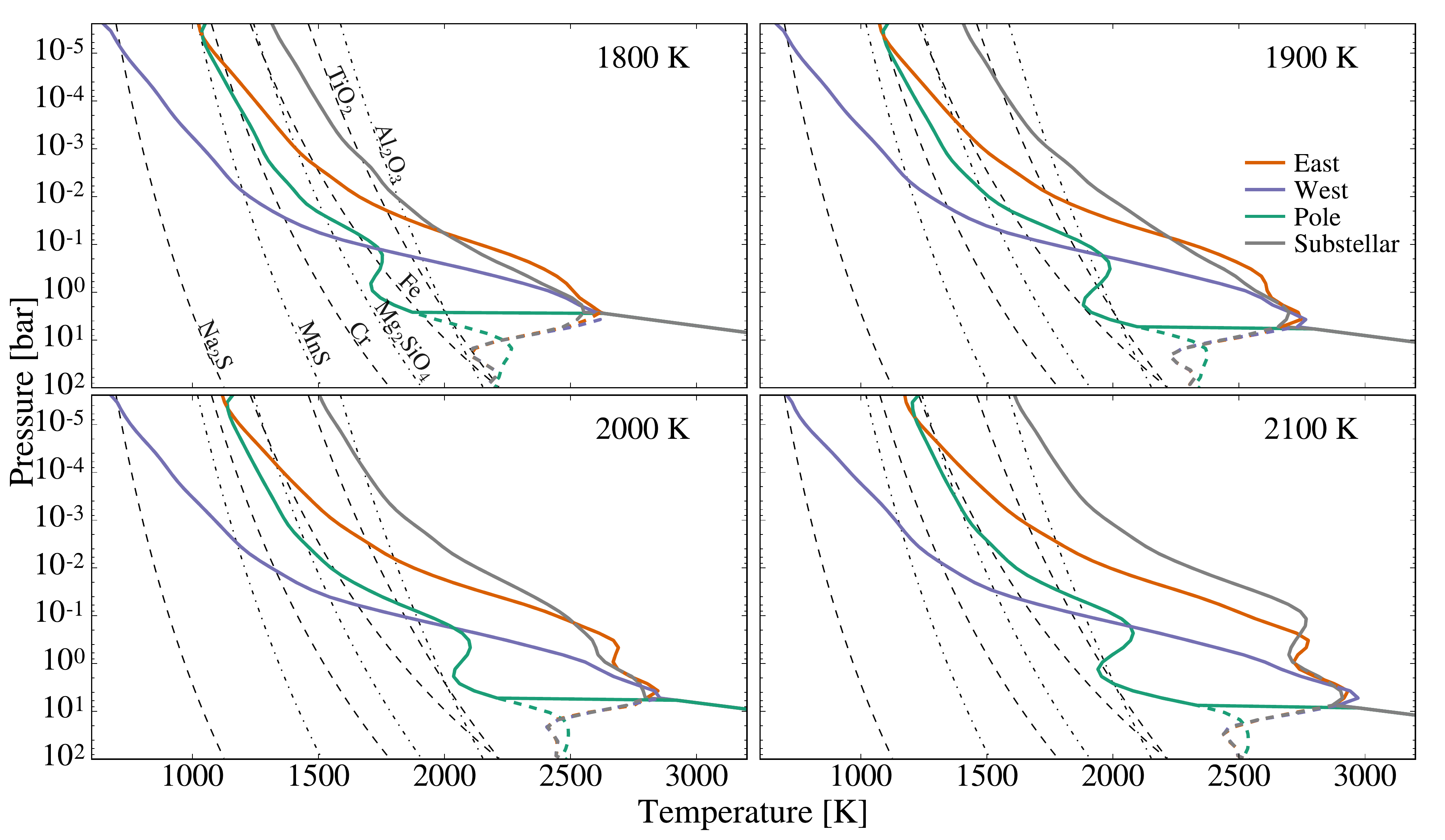}
\caption{Pressure temperature profiles at the east limb, west limb, and poles for four planets with equilibrium temperatures ranging from 1800 - 2100 K (solid lines). The original interior pressure temperature profiles from the SPARC/MITgcm \citep{2016ApJ...828...22P} are shown in dashed, colored lines. The temperature profiles used in our modeling converge to an internal adiabat at a few bar. The dashed black lines indicate the condensation curves for the different species that we consider in our modeling. If a temperature profile crosses a condensation curve then cloud formation may occur if there are no other significant barriers to nucleation and growth. }
\label{pt_profs}
\end{figure*}

In our model setup, gaseous species initially diffuse from the lower atmosphere until they reach a point in the atmosphere where the gas becomes supersaturated and cloud formation occurs via nucleation. In this model, the cloud species with the lowest surface tension (TiO$_2$) homogeneously nucleates and forms clouds. Once these cloud particles grow to a large enough size such that curvature effects no longer prevent heterogeneous nucleation, they become cloud condensation nuclei for other cloud species which are treated separately after nucleation occurs. These species will continue to grow through either condensation or coagulation. This growth is inhibited primarily by gravitational settling which causes cloud particles to fall to hotter regions of the atmosphere and quickly evaporate. For specific values used in our model setup see Table \ref{mps}. 

\section{Simulation Cases}\label{sim_cases}
Ideally, we would fully couple our microphysical cloud model to a 3D general circulation model (GCM) \citep[e.g.,][]{2018A&A...615A..97L}. However, these simulations are currently computationally expensive, such that running a large grid of models is not yet feasible, and dependent on initial conditions. As a first approximation, we use output from a GCM to determine the temperature structure of the atmosphere and then compute cloud properties with our 1D model at specific locations along the terminator, namely the east limb, west limb, and poles. A schematic of our model setup is shown in Figure \ref{model_setup}. This approach is similar to that in \citet{2019A&A...626A.133H,2019arXiv190608127H}, however, we model a grid of planets instead of a detailed study of an individual planet and we particularly focus on the cloud properties along the atmospheric limbs accessible to transmission observations.

\subsection{Pressure/Temperature Profiles}
We consider a grid of four hot Jupiters that range in equilibrium temperature from 1800 - 2100 K derived from the SPARC/MITgcm as presented in \citet{2016ApJ...828...22P}. We utilize the GCM derived temperature profiles as shown in Figure \ref{pt_profs} down to a pressure where interior models indicate that the temperature structure becomes adiabatic \citep{2019arXiv190707777T,Gao2019} - typically around the few bar level for planets with such high equilibrium temperatures. To create these temperature profiles we take the radiative-convective boundaries for each equilibrium temperature from \citet{2019arXiv190707777T} (see their Figure 4) which correspond to internal temperatures of $\sim$ 700 K for the planets in our sample. Below the radiative-convective boundary we assume that the temperature follows an adiabatic gradient for molecular hydrogen from \citet{2015A&A...574A..35P} (see their Equation 13). At this point the atmosphere is optically thick such that assumptions about the deep atmosphere will not change the resulting gas opacities. The temperature structure of hot Jupiter interiors is highly uncertain, however, and a variety of internal structures are likely necessary to explain the observed diversity in radii \citep[e.g.,][]{2014arXiv1405.3752G,2017ApJ...844...94K}. While assumptions about the planetary interior structure may alter the inferred cloud properties through the presence (or lack) of a deep cold trap \citep[see][]{2018ApJ...860...18P}, the results presented in this work are insensitive to the assumed interior temperature structure due to the lack of efficient deep cold-traps at all planetary locations sampled in this study.

The temporally averaged limb temperature profiles are taken from the GCM at longitudes of -90$^{\circ}$ (west limb) and 90$^{\circ}$ (east limb) and are latitudinally averaged. The temporally averaged polar profile is sampled at a latitude of 90$^{\circ}$. These temperature profiles thus differ from those presented in \citet{2018ApJ...860...18P}. We note that these profiles do not include TiO or VO gas phase opacities which could lead to a temperature inversion and potentially alter the temperature structures across the globe. This increase in temperature would generally lower the cloud formation efficiency by lowering the atmospheric supersaturation of the forming cloud species. the presence of gas phase TiO in atmospheres with equilibrium temperatures presented in this work, however, is highly uncertain. The resulting temperature profiles used for the modeling in this work for each planet in our grid are shown in Figure \ref{pt_profs}. 

We sample each atmosphere at the east limb, west limb, and polar region and calculate cloud properties. We sample the limbs and poles in particular as we are interested in the planetary properties as viewed in a transmission viewing geometry. Every model planet in our sample has an east limb temperature structure that is significantly hotter than the west limb and the pole at all pressures lower than 1 bar (see Figure \ref{pt_profs}). While all of the planets in our grid are relatively hot such that the efficiency of their heat redistribution is low, they are not uncommon in the known sample of hot Jupiters \citep[e.g.,][]{2018AJ....156..122M,2017ApJ...847L..22F,2018AJ....155..156T,2016Natur.529...59S,2019AJ....157...68B,2018AJ....155...83Z,2013ApJ...776L..25D,2015ApJ...802...51H,2015ApJ...804...94W,2015AJ....150..112S,2014ApJ...785..148R,2015ApJ...811..122W,2017A&A...608A..26M}.

\subsection{Atmospheric Vertical Mixing}\label{kzz_section}
The amount of vertical mixing in an atmosphere regulates the cloud formation process through delivering a fresh supply of condensible volatiles to the upper atmosphere where cloud formation can occur. In the supply-limited regime of cloud formation modeled in this work, the higher the vertical mixing in the atmosphere, the more cloud formation occurs \citep{2018ApJ...860...18P}. 

 \begin{figure}[tbp]
\epsscale{1.2}
\plotone{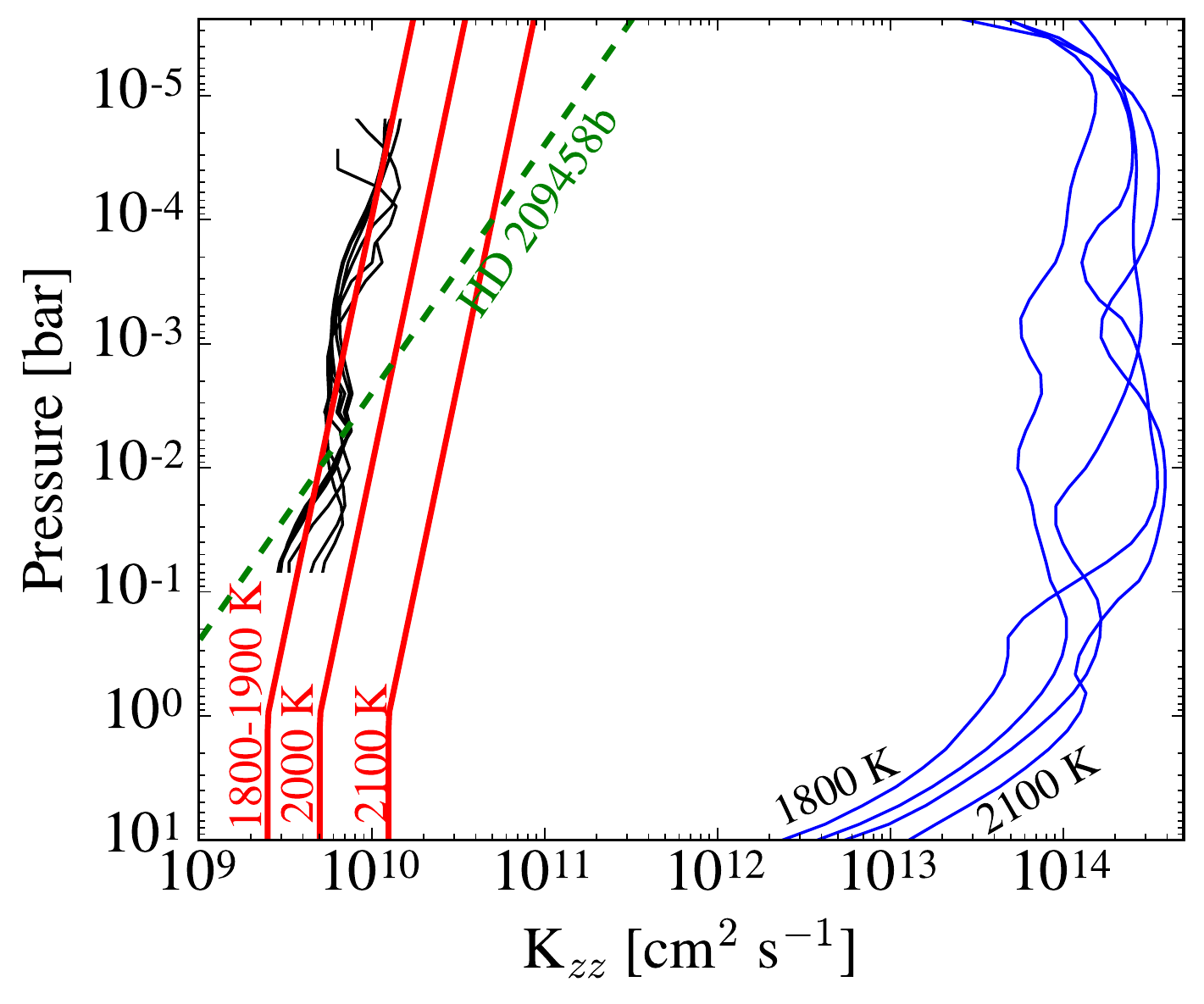}
\caption{Globally averaged K$_{zz}$ profiles used in this work (red lines) fit to the transport of tracers with sizes 0.1-100 $\mu$m (black lines, for the 1800 K case) with a power-law in the upper atmosphere and a constant value below 1 bar. The previous tracer derived K$_{zz}$ profile for the cooler hot Jupiter, HD 209458b, is also shown (green line) as well as the roughly approximated K$_{zz}$ values derived for each planet (blue lines) as a global area-weighted root mean square of the vertical velocity times the vertical scale height. For every planet, the tracer-derived vertical mixing profile is roughly three orders-of-magnitude less than the approximate profile derived from the vertical wind speed.}
\label{kzz}
\end{figure}

We use globally averaged vertical mixing profiles for each planet derived from general circulation models that include tracer transport. We parameterize all vertical motions in the planetary atmosphere using eddy diffusion, controlled by a diffusive term known as K$_{zz}$. These profiles are calculated using time-averaged SPARC/MITgcm simulations for a highly irradiated Jupiter-sized planet without the hot internal adiabat below 1 bar \citep{parmentier2016transitions,parmentier2019}. The method used to derive these K$_{zz}$ profiles follows \citet{2013A&A...558A..91P} (see their eq. 23) and \citet{2018ApJ...866....1Z,2018ApJ...866....2Z} and depends on the tracer gradient, which can be both positive or negative. For each planet, we fit a power-law to the tracer derived K$_{zz}$ values, as shown in Figure \ref{kzz}, and assume a constant value below 1 bar as the GCM derived K$_\text{zz}$ is inaccurate at higher pressures due to integration time. We note that the mixing profile below the 1 bar level may be able to be approximated convectively. However, in our tests the steady state cloud properties are insensitive to increasing or decreasing the K$_\text{zz}$ below 1 bar by a factor of 5.

For all of the planets in our sample, the vertical mixing operates roughly the same for particles that range in size from 0.1-100 $\mu$m. There is also less of a dependence of mixing strength on atmospheric pressure in the upper atmosphere as was derived for the cooler hot jupiter, HD 209458b \citep{2013A&A...558A..91P}. Furthermore, the value of the vertical mixing is smaller by roughly 3 orders-of-magnitude compared to the global root mean square of the vertical velocity multiplied by the vertical scale height as shown in Figure \ref{kzz}, a common estimate of K$_{zz}$ in the literature. 

The K$_{zz}$ profile that best describes the planets in our sample with T$_\text{eq}$ of 1800 and 1900 K is:

\begin{equation}\label{kzz_prof}
    \begin{array}{@{} r @{} c @{} l @{} }
&K_{zz} &{}=\displaystyle
\begin{cases}
2.5\times 10^9 \text{cm}^2 \text{s}^{-1}/P_\text{bar}^{0.15} & P < 1\text{ bar},\\
2.5\times 10^9 \text{cm}^2 \text{s}^{-1} &  P > 1\text{ bar}.
\end{cases}
\end{array}
\end{equation}

For planets with T$_\text{eq}$ = 2000 and 2100 K, the Kzz profiles are factors of 2 and 5 larger than that in Equation \eqref{kzz_prof}. The globally averaged vertical mixing in our modeling thus slightly increases with increased equilibrium temperature in this model range as predicted in \citet{2019ApJ...881..152K}.

\subsection{Choice of Temperature Range}

Our grid has a temperature range of 1800 - 2100 K. We chose this range as this regime may correspond to a maximum in limb cloud inhomogeneity. While most hot Jupiters have significant temperature gradients from east to west that could very well lead to inhomogeneous cloudiness, cloud cover may be able to efficiently homogenize between the two limbs if both the atmospheric circulation from the cooler west hemisphere to the hotter east hemisphere is efficient and clouds are able to survive crossing the dayside of the planet or clouds are able to form efficiently on the dayside itself.

There are two possible dominant circulation patterns on hot Jupiters that work to equilibriate insolation gradients \citep{showman-etal-2013}. The first circulation pattern is jet dominated and is characterized by an efficient superrotating wind across the planetary equator from west to east where the efficiency with which this flow structure equilibriates the planetary temperature decreases with increasing equilibrium temperature \citep[e.g.,][]{2016ApJ...821...16K}. The second pattern is eddy dominated and is characterized by large scale flows from the dayside to both the nightside and the two limbs. The GCM models presented in this work do not contain additional drag to the numerical one \citep{2018ApJ...853..133K} such that they are in the jet-dominated regime. At these temperatures, however, magnetic effects or others might drag the winds and drive the flow towards the eddy dominated regime which may homogenize the temperature structure at the limbs \citep{showman-etal-2013}. The strength of the drag in these atmospheres depends on planetary properties, such as the magnetic field strength, such that there may be planets with drag and planets without. We thus assume that the planets presented in this work have flows that are dominated by west to east advection such that the west limb can only directly advect material to the east limb via the superrotating equatorial wind. In some cases, at high, cooler, low-area, latitudes on the dayside there may still be a connection between the east and west terminator regions via contra-rotating flows sometimes seen at higher latitudes in GCM models. The models presented in this work, however, do not exhibit such a flow pattern. 

\begin{figure}[tbp]
\epsscale{1}
\plotone{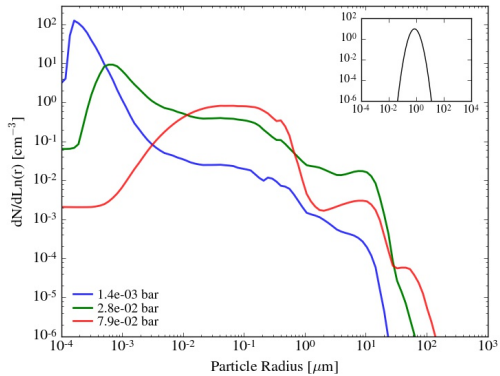}
\caption{Combined particle size distributions for all cloud species at various atmospheric pressure levels for an 1800 K hot jupiter at the west limb. These size distributions are not log-normal and exhibit distinct bumps due to the different formation modes (i.e. nucleation mode vs. growth mode) of different cloud species. A log-normal size distribution is shown for reference. }
\label{sd_single}
\end{figure}

 \begin{figure*}[tbp]
 \epsscale{1.2}
\plotone{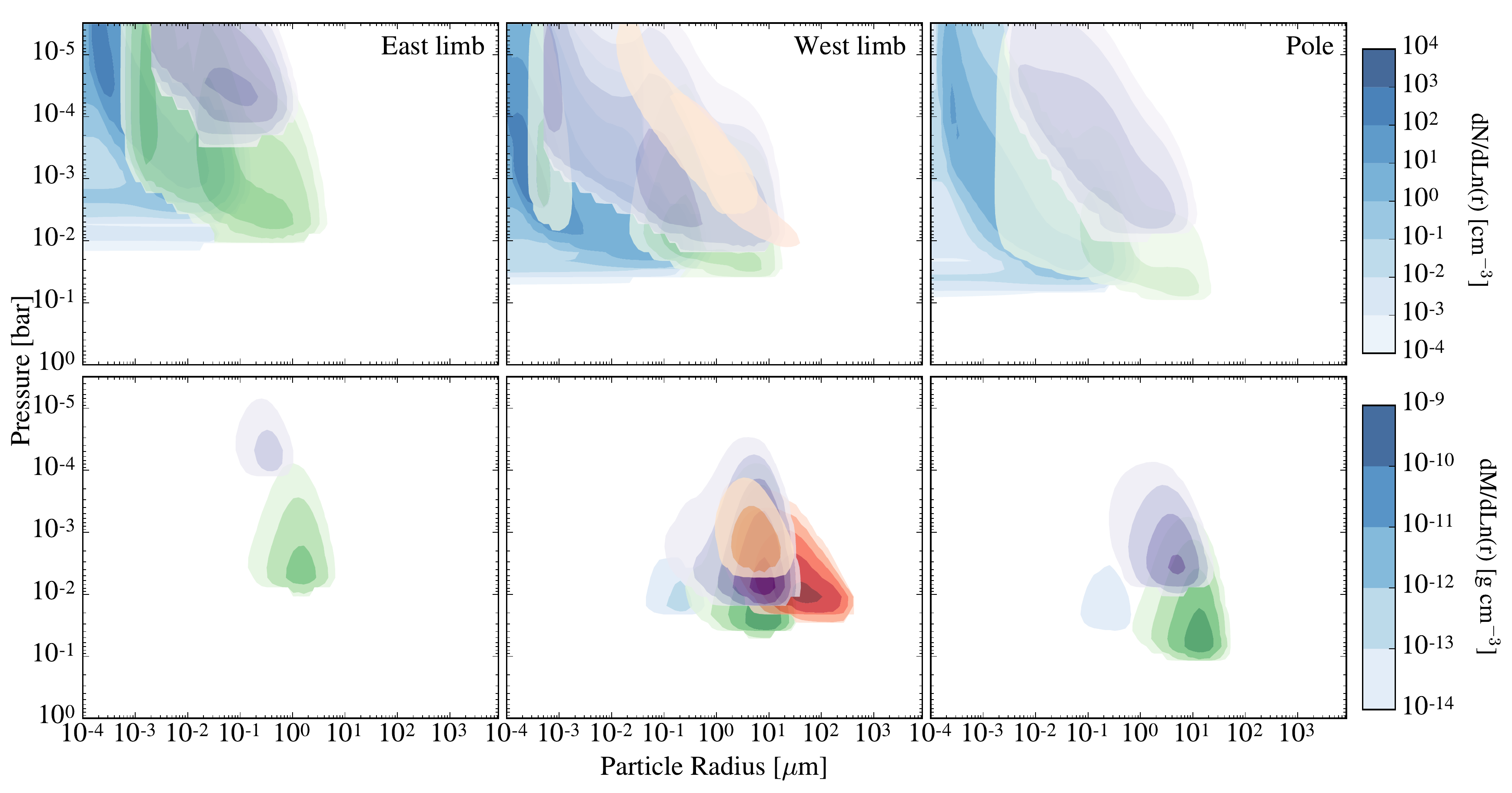}
\caption{Particle number (top) and mass density (bottom) distributions at the east limb (left) west limb (middle) and pole (right) for a hot jupiter with an equilibrium temperature of 2000 K. The cloud species shown are TiO$_2$ (blue), Mg$_2$SiO$_4$ (purple), Al$_2$O$_3$ (green), Fe (red), and Cr (orange).}
\label{2000_sd}
\end{figure*}

For hot Jupiters with equilibrium temperatures larger than 1800 K, it is possible that no species of cloud forms on the bulk of the dayside as it is too hot for titanium clouds, the most likely cloud condensation nuclei \citep{2018A&A...614A.126L}, to form and serve as nucleation sites for cloud species with higher condensation temperatures. If planets in this temperature range do still form superrotating equatorial jets, the clouds that form on the west limb will rapidly evaporate when advected across the dayside as the timescale for evaporation is very short, on the order of seconds or less when thermodynamically favorable \citep{2018ApJ...860...18P} while the time to cross the dayside is $\sim R_p/v_\text{advect} \sim 10^5$ seconds. Thus the cloud distribution on the east limb is likely isolated from the other more efficient cloud forming regions. We note that it may be possible that clouds on the east limb influence the west limb after being transported across the nightside as has been suggested by previous GCM modeling \citep{2018A&A...615A..97L}. At high equilibrium temperatures, however, the west limb cloud structures may dominate observationally as the cloud formation timescales at the cooler west limb and advection timescale are similar \citep{2018ApJ...860...18P} meaning that clouds from the east limb may evolve to have west limb cloud properties once they have been advected. Thus, for hot Jupiters with T$_\text{eq} > 1800$ K, the east and west limbs on hot Jupiters might represent the most extreme case of inhomogeneous cloud cover. We further choose a maximum equilbrium temperature in our model grid of T$_\text{eq}$ of 2100 K as hotter planets may reside in a different regime due to increased magnetohydrodynamic effects. 

\subsection{Choice of Microphysical Parameters}\label{minmax}
There are two key microphysical parameters that regulate heterogeneous nucleation that are not currently well-constrained: the contact angle and the desorption energy (See Appendix \ref{props}). We discuss the sensitivity of our results to these microphysical parameters in Section \ref{micro_sensitivity}. 

In this work, we approximate each species' desorption energy as half of its calculated latent heat of vaporization. This approximation has previously been used to estimate the desorption energy of water and other condensible species and may thus be the closest estimate we have for these values without detailed laboratory experiments \citep[e.g.,][]{doi:10.1098/rsta.2002.1137,2013ctmc.book....3B,Hyunho2016}. These values are given in the appendix in Table \ref{desorption}. For the contact angle we choose the minimum theoretically motivated value, leading to a maximum in cloud formation, such that $\cos{\theta_c} = \sigma_C/\sigma_x$, where $\theta_c$ is the contact angle, $\sigma_C$ is the surface energy of the cloud condensation nuclei and $\sigma_x$ is the surface energy of the condensible species.

\section{Cloud Properties and Particle Size Distributions}\label{clouds}

For each location on the planet we calculate cloud particle size distributions from first principles as a function of atmospheric depth. These particle size distributions change with cloud composition and atmospheric pressure level (Figure \ref{sd_single}) and are typically broad and irregular in shape. The shape of these particle size distributions gives rise to atmospheric features and changes the shape of the transmission spectra as described in Section \ref{trans}. 

 \begin{figure}[tbp]
\epsscale{1}
\includegraphics[width=\columnwidth]{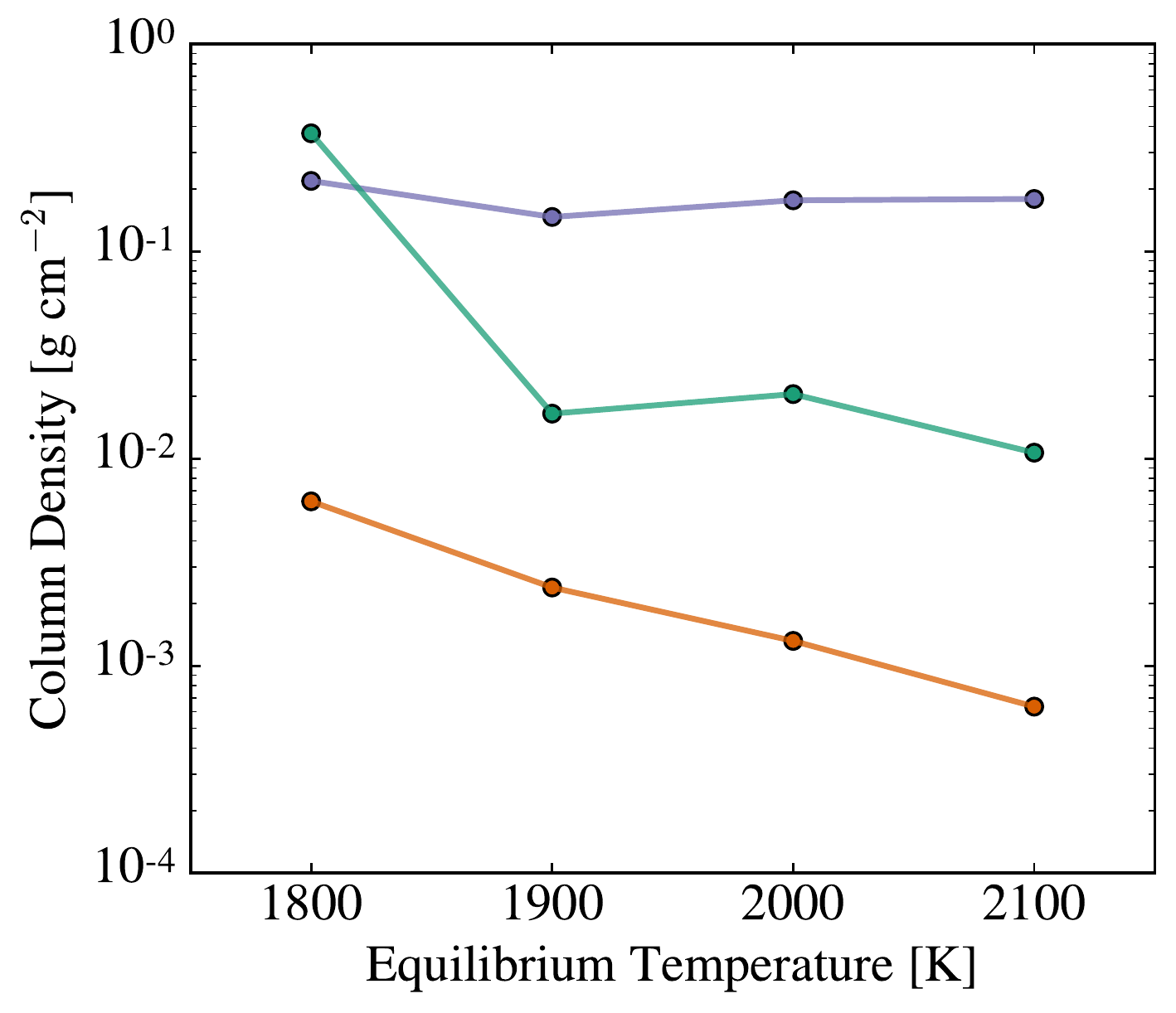}
\caption{The column-integrated condensed mass density at the west limb (purple) exceeds that at the east limb (orange) for all equilibrium temperatures, and those at the pole (green) for all but the coolest equilibrium temperature. The planet with T$_\text{eq} = 1800$ K has a more mass at the pole than the west limb, though the majority of the mass is present in the deep atmosphere and does not contribute to the observed spectra. }
\label{masstrend} 
\end{figure}

\begin{figure*}[tbp]
\epsscale{1.15}
\plottwo{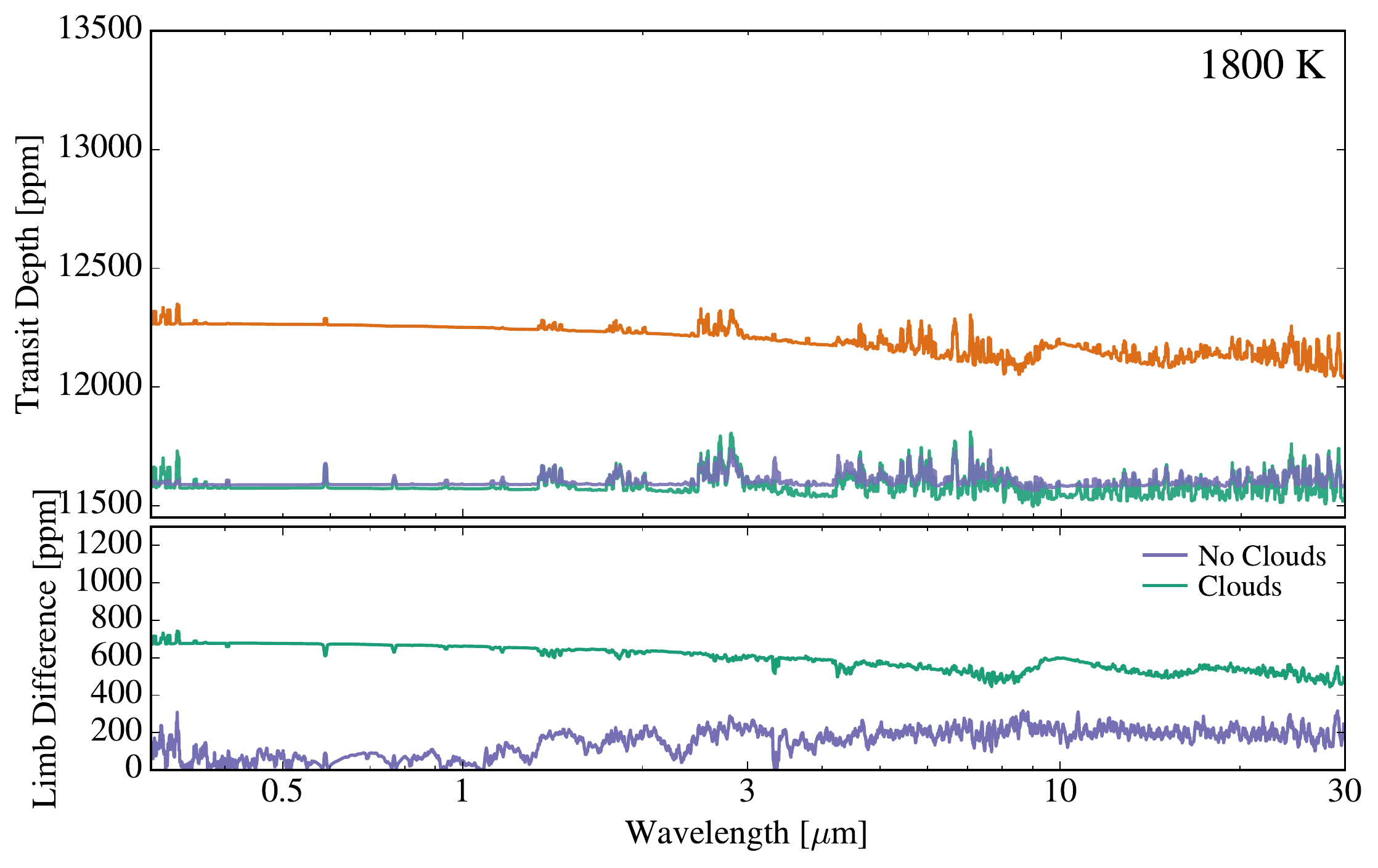}{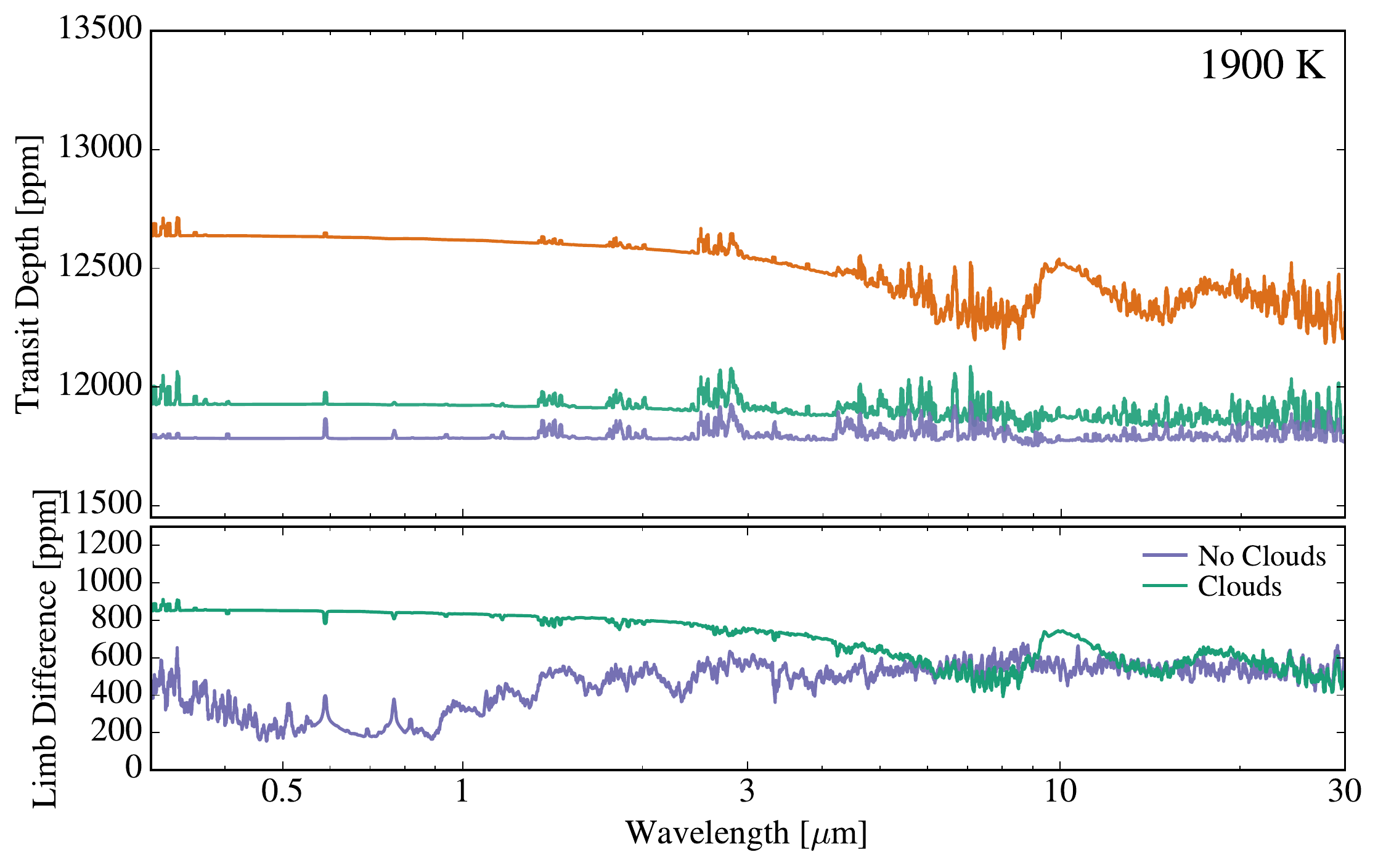}
\plottwo{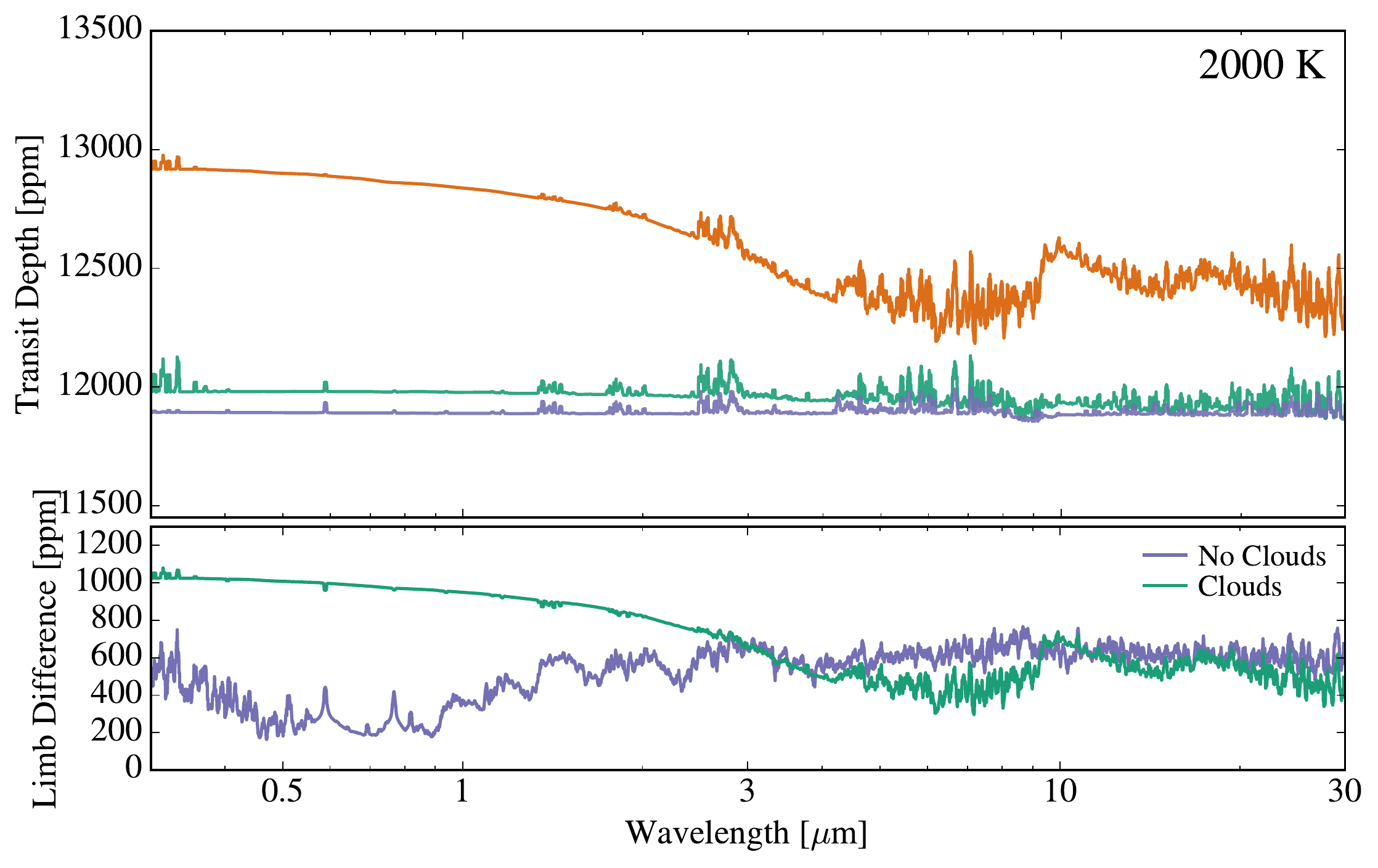}{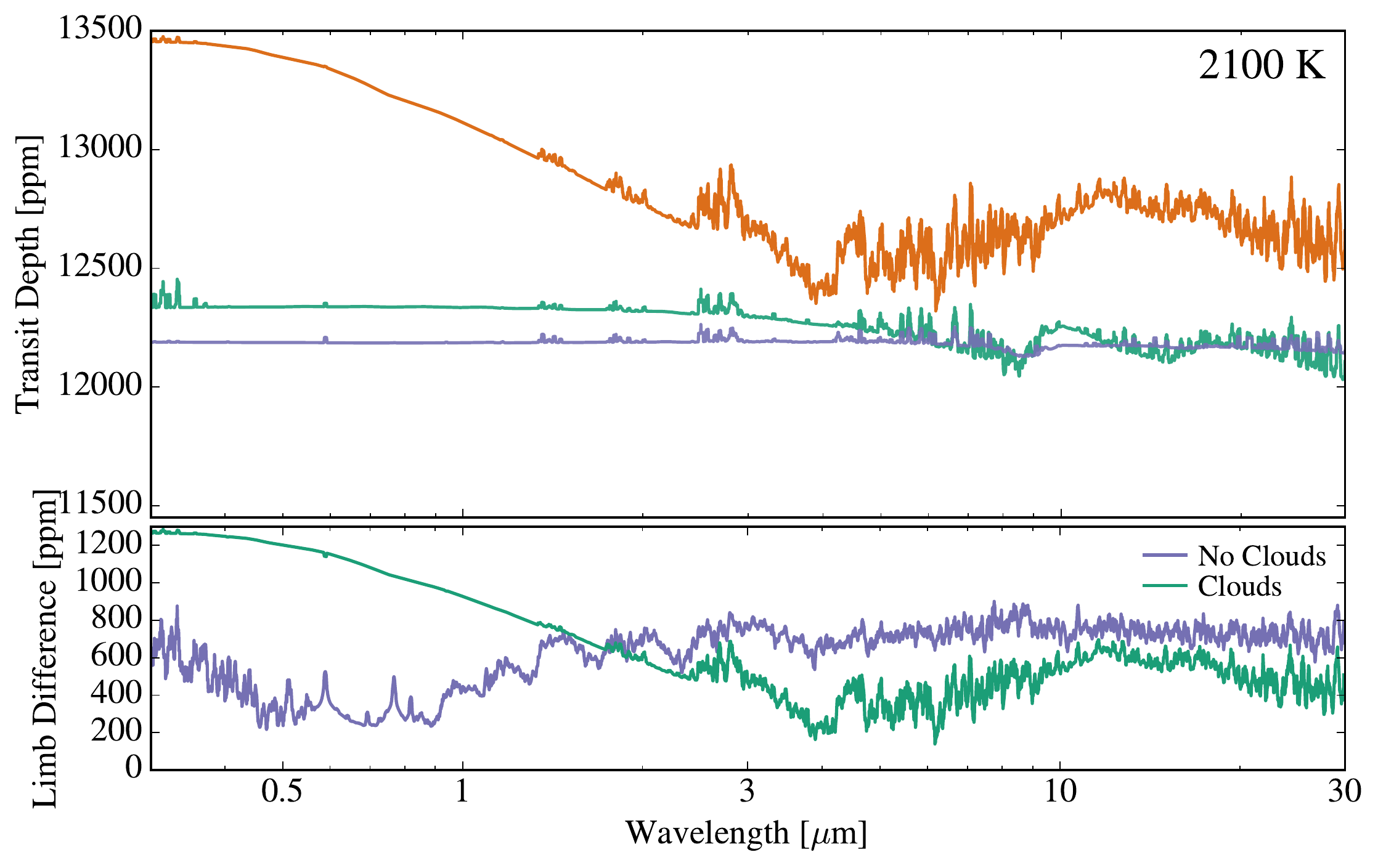}
\caption{Transmission spectra (top half of each plot) for the east limbs (orange), west limbs (purple), and poles (green) at different equilibrium temperatures, and the difference between the limbs in the cloudy (green) and clear (purple) cases.}
\label{max_trans}
\end{figure*}

The full two-dimensional cloud particle size distributions are shown for a nominal case of T$_\text{eq} = 2000$ K in Figure \ref{2000_sd} for the three locations sampled in our grid. The cloud base of the different cloud species considered are often distinct from each other as they become supersaturated at different pressure levels in the atmosphere. This is particularly true at the east limb. Only the polar region of the T$_\text{eq}$ = 1800 K exhibits a deep cold trap \citep[see][]{2018ApJ...860...18P}. This cold trap is inefficient, however, as there is significant cloud formation in the upper atmosphere that contributes to the observed atmospheric opacity.

Interestingly, MnS clouds do not form. This is due to the relatively high surface energy of MnS which exponentially regulates nucleation efficiency and is roughly 10\% larger than Fe, the species with the next largest surface energy. The degree of MnS supersaturation is also low compared to the nucleation barrier stemming from its high surface energy (see Figure \ref{pt_profs}) such that cloud formation does not occur. Iron, however, is able to form despite its high surface energy but is only able to do so at the west limb where it is significantly supersaturated. This illustrates that the condensation curve alone does not definitively describe when cloud formation will occur and detailed non-equilibrium microphysical studies are important when interpreting observations. 

In all of the simulated cases, the west limb has over an order of magnitude higher condensed mass density than the east limb as shown in Figure \ref{masstrend}. Furthermore, the polar region of each planet has lower cloud mass density than the west limb in the upper regions of the atmosphere that contribute to the observed spectra.

\section{Transmission Spectra}\label{trans}
We modify Exo-Transmit \citep{2017PASP..129d4402K} to consider the opacity from the fully resolved cloud particle size distributions calculated for these model atmospheres. We treat the cloud particles as Mie spheres and calculate the particle extinction cross sections using \textsc{MIEX} \citep{2004ASPRv..12....1V}. Complex indices of refraction for Cr are taken from the compilation of \citet{2012ApJ...756..172M}, those of TiO$_2$ are compiled in \citet{2003ApJS..149..437P,2011A&A...526A..68Z} and those of Fe, Mg$_2$SiO$_4$ (crystalline), and Al$_2$O$_3$ are taken from the compilation in \citet{2015A&A...573A.122W}. In these calculations, we treat each cloud independently. While all cloud species other than TiO$_2$ are inhomogeneous in composition (with a TiO$_2$ core and a mantle of the primary condensible species) the optical properties are treated as that of the primary condensible species, an approximation that does not change the resultant spectral features in our results\footnote{An test analysis of the spectra using \textsc{pymiecoated}, which calculates the optical properties of layered mie spheres, produces the same results.}. We calculate the abundance of the gaseous species assuming equilibrium chemistry with solar abundances including the rainout of condensible species. We do not include gaseous TiO or VO, consistent with the pressure/temperature profiles presented in this work, as the presence of these strong atmospheric absorbers in this range of equilibrium temperatures is uncertain. While TiO may well be present on the cloud-free daysides of these planets, we expect that much of the atmospheric TiO on the limbs of planets in this regime may be located in condensible species. All of the presented transmission spectra in this work have a binned resolution of R = 10$^2$.

For every planet in our grid, the transmission spectra at the east and west limbs are significantly different (Figure \ref{max_trans}). In particular, the transmission spectra of the east limb appear more clear and have noticeable molecular features at longer wavelengths. The west limb, however, appears significantly more cloudy with much more subdued molecular features. There is also a significant continuum difference between the two limbs such that the difference in spectra between the limbs can be as much as 1000 parts per million for a broad range of wavelengths. The transmission spectra for the polar region is similar to the west limb, though there are typically more spectral features observable at longer wavelengths at the polar regions.

The difference in transmission spectra between different limb locations arises from differences in the atmospheric thermal structure which significantly alters both the cloud opacity and the total gaseous opacity. In particular, the local cloud properties are different at atmospheric locations with different temperature structures such that clouds form at different heights in the atmosphere with potentially different compositions as shown in Section \ref{clouds}. We discuss the specific effects of cloud properties on the transmission spectra in the following sections. 

\subsection{Observed Cloud Height}

\begin{figure}[tbp] 
\epsscale{1.15}
\plotone{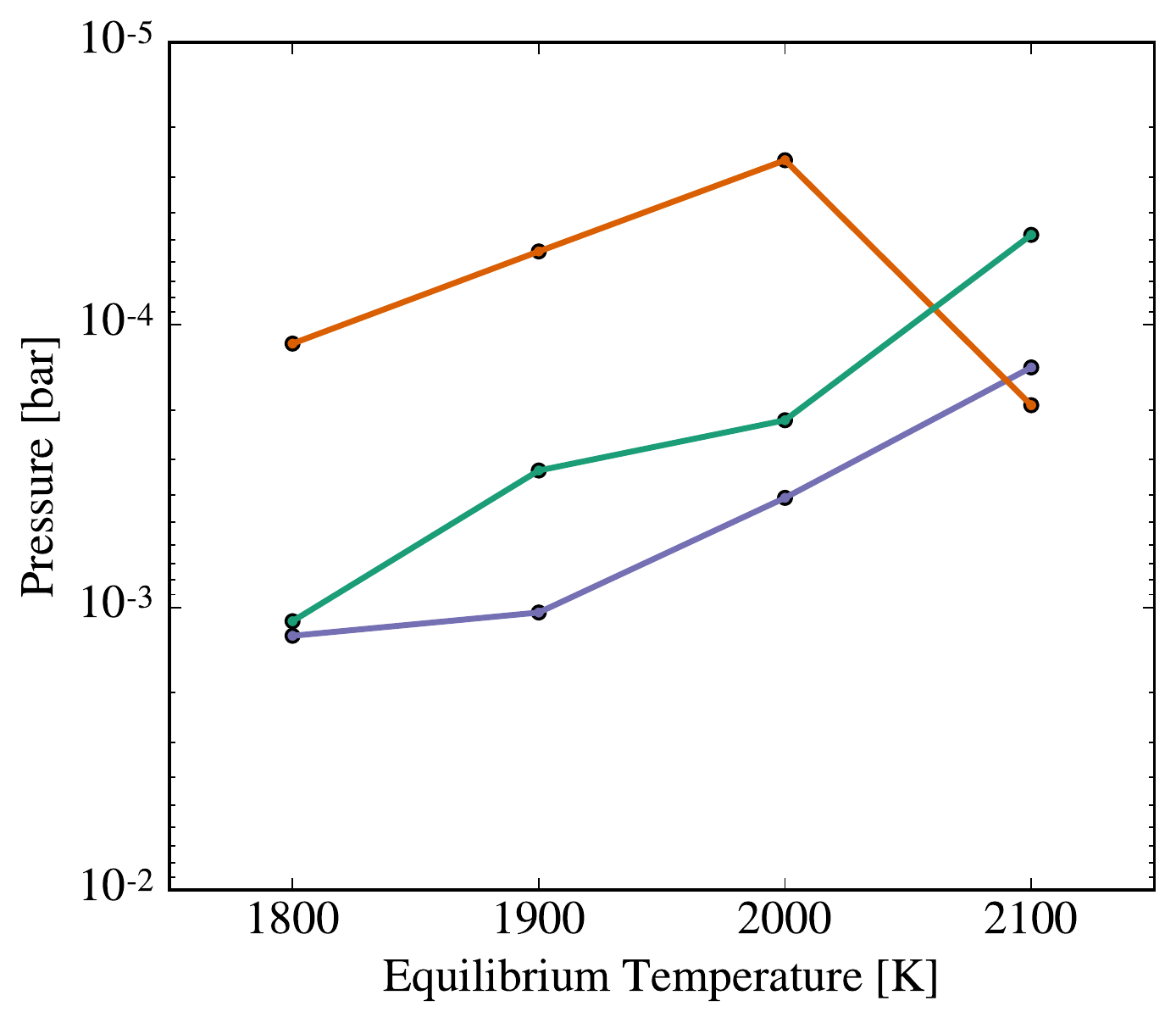}
\caption{The opaque cloud layer at 1.2 $\mu$m at the west limb (purple), east limb (orange), and pole (green). }
\label{cloud_height}
\end{figure}

\begin{figure}[tbp]
\epsscale{1.15}
\plotone{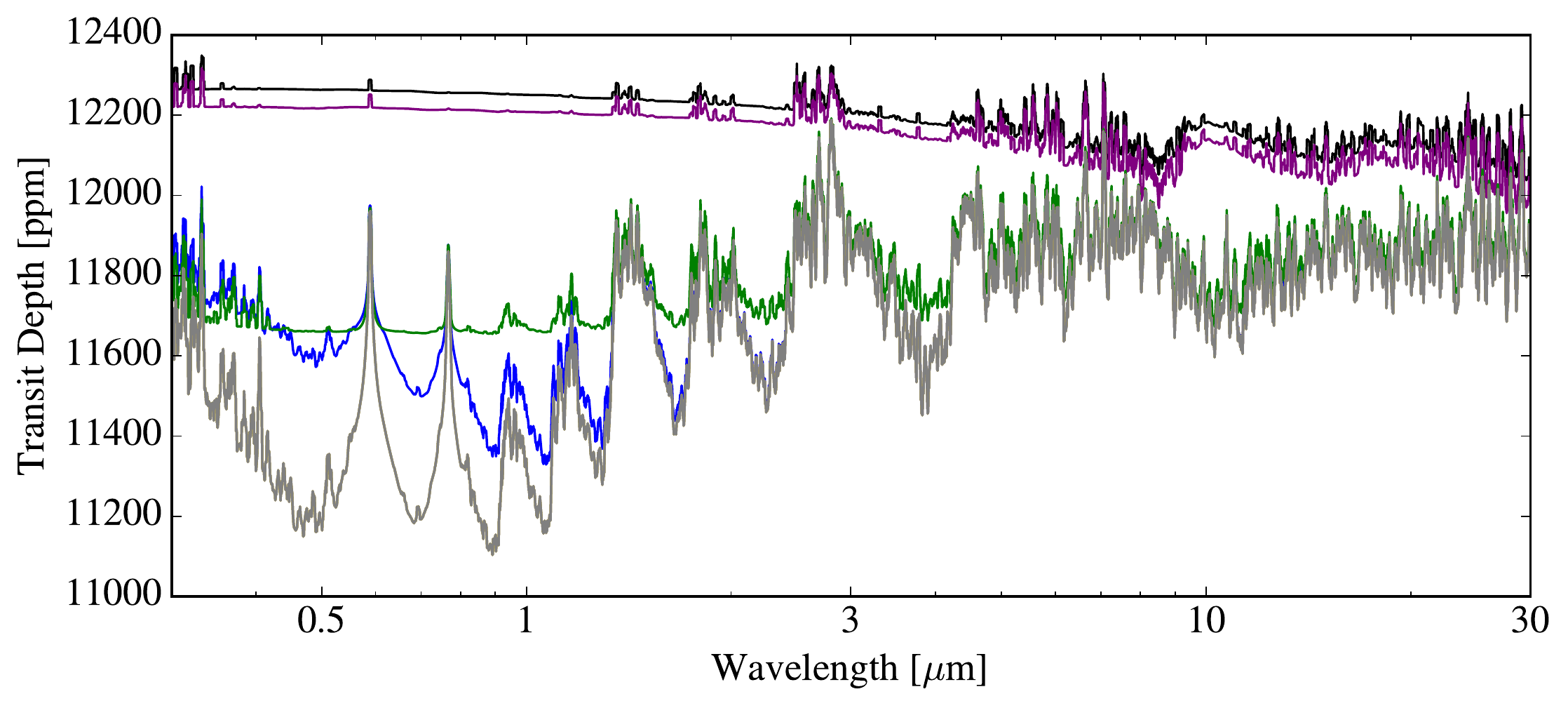}
\plotone{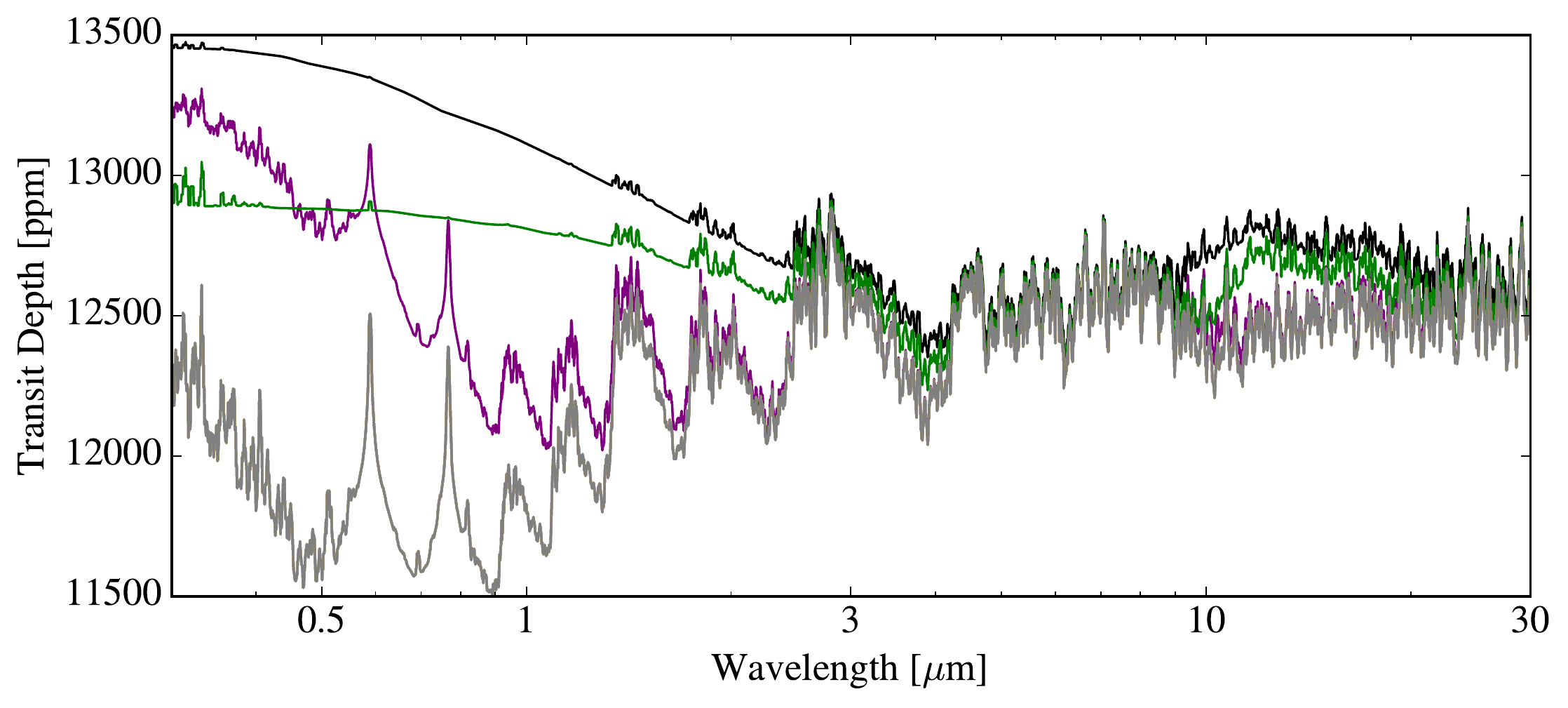}
\caption{The contribution to the transmission spectra (black lines) from each cloud species for the 1800 K east limb (top) and the 2100 K east limb (bottom). Clear spectra for these planets are shown in gray. The cloud opacity is primarily dominated by silicate clouds (purple line) at all wavelengths except for the hottest regions of the hottest planets where aluminum clouds (green lines) play an increasingly significant role in shaping the spectrum. Titanium clouds (blue lines) primarily contribute to the total cloud opacity at short wavelengths. }
\label{species_contribute}
\end{figure}

While the west limbs appear more cloudy when observed across a broad wavelength range due to significantly flattened spectral features (see Figure \ref{max_trans}), the east limb can appear more cloudy than the west limb, particularly at wavelengths shorter than $\sim$ 2 $\mu$m. This occurs because, while there is less total cloud mass on the east limb, the altitude where the relevant cloud species are supersaturated (the cloud base) is higher in the atmosphere (see Figure \ref{pt_profs}). Cloud formation is typically the most efficient near the cloud base \citep{2018ApJ...860...18P}. Thus, the higher cloud base can give rise to clouds that are opaque higher in the atmosphere with relatively low total cloud mass. 

\begin{figure*}[tbp]
\epsscale{1.}
\plotone{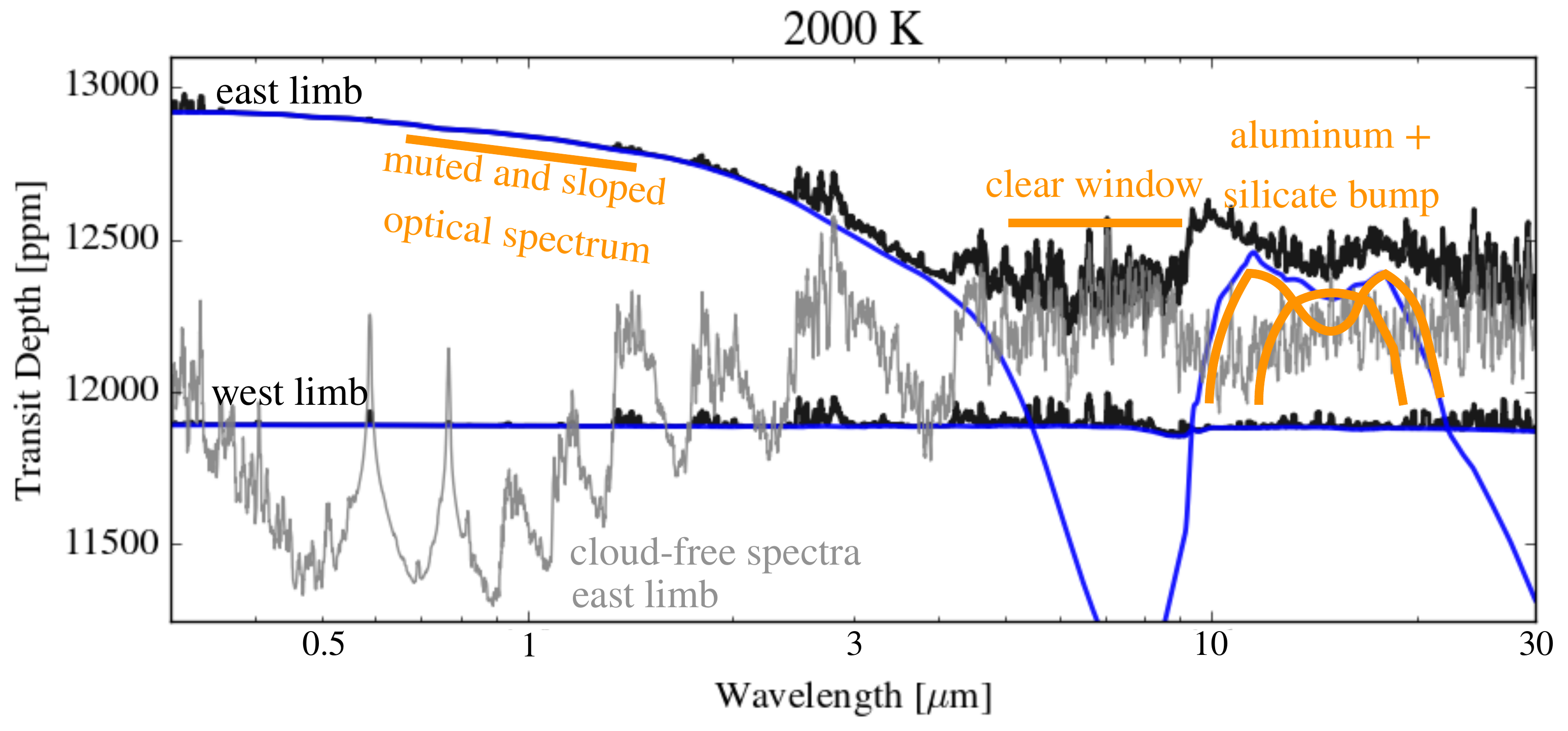}
\caption{Transmission spectra (black lines) for a hot jupiter with an equilibrium temperature of 2000 K at the east and west limbs. The blue lines are the opacity continuum from clouds. The cloud-free transmission spectrum at the east limb is shown in gray. At the west limbs, clouds dominate the spectra at all wavelengths. At the east limb, clouds contribute to muted transmission features at short wavelengths and a sloped optical spectrum. There is a relatively clear window at $\sim$ 5-9 microns and enhanced silicate and aluminum cloud opacity from 10-20 microns.}
\label{features}
\end{figure*}

The pressure level where clouds become opaque in transmission (the cloud height) at 1.2 $\mu$m is shown in Figure \ref{cloud_height} where, for both cloudiness cases at all equilibrium temperatures cooler than 2100 K, the cloud height is higher in the atmosphere along the east limb. Thus, while the east limb transmission spectra appear relatively clear at longer wavelengths where the clouds are less opaque, the features in transmission at short wavelength are often more damped in the hotter regions of the atmosphere with lower cloud mass.

\subsection{The Dominant Cloud Species}\label{dominant_clouds}
We analyze each cloud species' contribution to the total opacity. As found in \citet{Gao2019}, in nearly all cases the cloud opacity is dominated by silicate clouds with only small contributions to the total opacity from titanium and/or aluminum clouds. For the hottest planets in our sample at the east limb, where silicate cloud formation becomes less efficient, however, aluminum clouds tend to increasingly dominate the observed spectra. This is particularly true for the hottest planet in our grid at the east limb, where aluminum clouds dominate the total cloud opacity and contribute to the Rayleigh-like slope in the optical. This is shown for two representative cases in Figure \ref{species_contribute}. 

Both chromium and iron clouds, which form on the west limbs of all of the planets in our grid, do not significantly impact the transmission spectra due to their relatively low number densities as shown in Figure \ref{2000_sd}. While both of these species are able to grow to large, massive sizes once heterogeneous nucleation onto titanium seeds has occurred (the first step in the cloud formation process for Cr and Fe clouds), the rate of heterogeneous nucleation is suppressed due to the relatively high surface tensions of these species. Thus, while both cloud species can form particles that grow to large sizes, the total number of cloud particles is small in comparison with the more abundant aluminum, titanium, and silicate clouds. It is possible, therefore that significant constituents of the total cloud mass on the west limbs, such as Fe and Cr clouds, can have no significant impact on the observed spectra. 

\subsection{Significant Cloud Transmission Features}\label{trans_features}

Our model spectra are strongly modulated by clouds as shown by muted spectral features, broad absorption features in the infrared, and sloped optical spectra. While clouds can give rise to a sloped optical spectrum, this can also arise due to stellar contamination and other effects \citep[e.g.,][]{2014A&A...568A..99O,2014ApJ...791...55M,2018arXiv180308708A}. However, the broad absorption features in the infrared in particular are clear, direct signatures of clouds.

Along the west limb for all of the planets, clouds act like gray absorbers and substantially damp the observed spectrum at all wavelengths as shown in Figure \ref{features} (see also Figure \ref{max_trans}). The damping of spectral features is a common outcome of the presence of clouds and occurs across a broad wavelength range when nucleation and condensation are efficient, as occurs on the west limbs. At the poles, we find similarly damped spectral features but begin to see a broad spectral signature of silicate clouds at $\sim$ 10 $\mu$m as predicted in \citet{2015A&A...573A.122W} and $\sim$ 20 $\mu$m \citep[i.e. 2100 K case in Figure \ref{max_trans}. Also see][]{2019MNRAS.487.2082L}. These broad spectral features are clear signatures of the presence of clouds and often have amplitudes on the order of $\sim$100s of ppm which should be feasible for detection using \textit{JWST} \citep{Venot2019,2017ApJ...850..121M}. 

The east limbs show the most significant cloud features as demonstrated in Figure \ref{features}. At the east limb, the less massive populations of small cloud particles high in the atmosphere give rise to a muted and sloped optical spectrum, with slopes that increase with increased temperature, as well as signatures of silicate (at $\sim$ 10 - 30 $\mu$m) and/or aluminum clouds (at $\sim$ 15 $\mu$m). Intriguingly, these hotter regions of the atmosphere have a cloud-free spectral window from $\sim$ 5 - 9 $\mu$m. These transmission features on the east limb are qualitatively similar to the mie slope and spectral window seen in the sub-neptune GJ 3470b \citep{2019NatAs.tmp..377B}. Thus, the marginally supersaturated regions of a planetary atmosphere, despite forming fewer clouds, frequently provide a more clear signature of both the species and properties of the clouds present in the atmosphere. 

\begin{figure*}[tbp]
\epsscale{1.15}
\plottwo{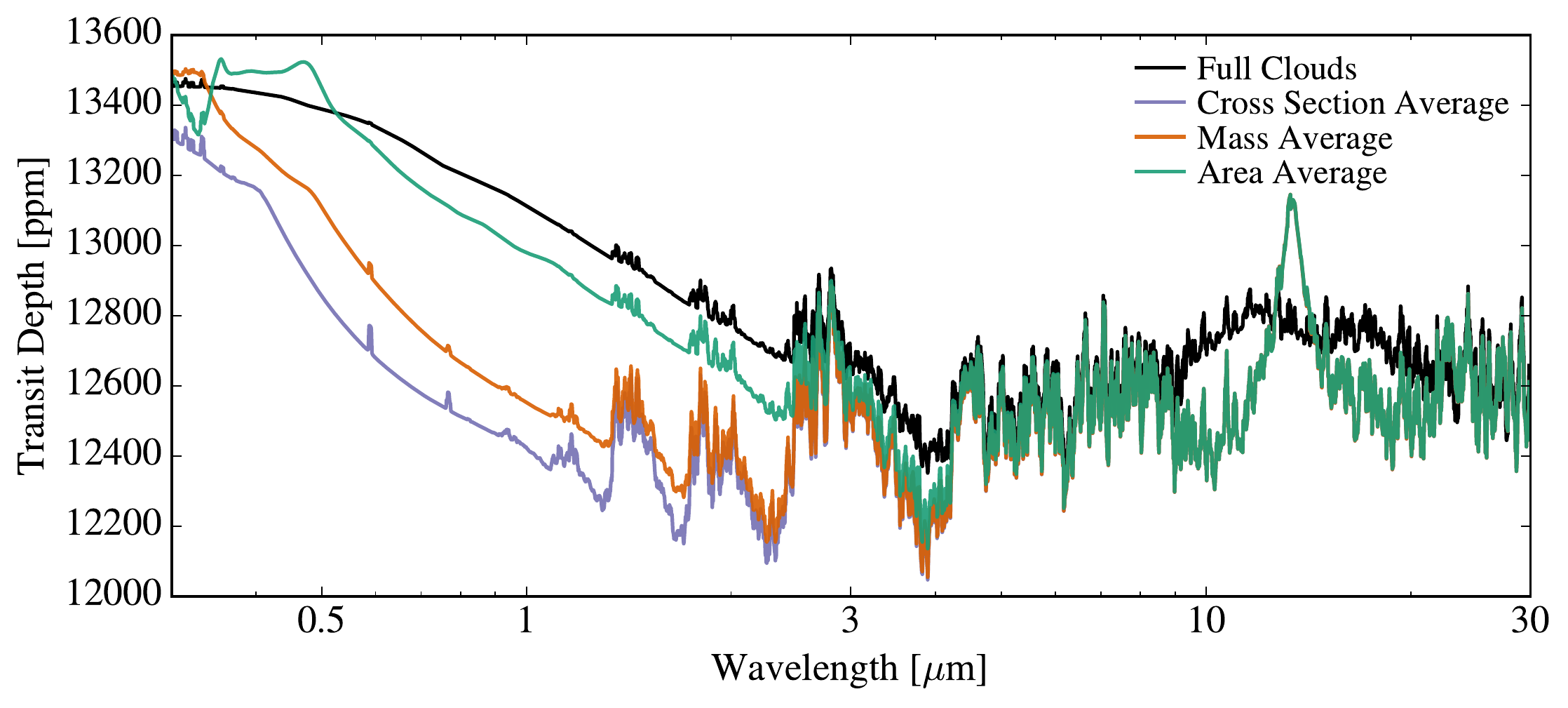}{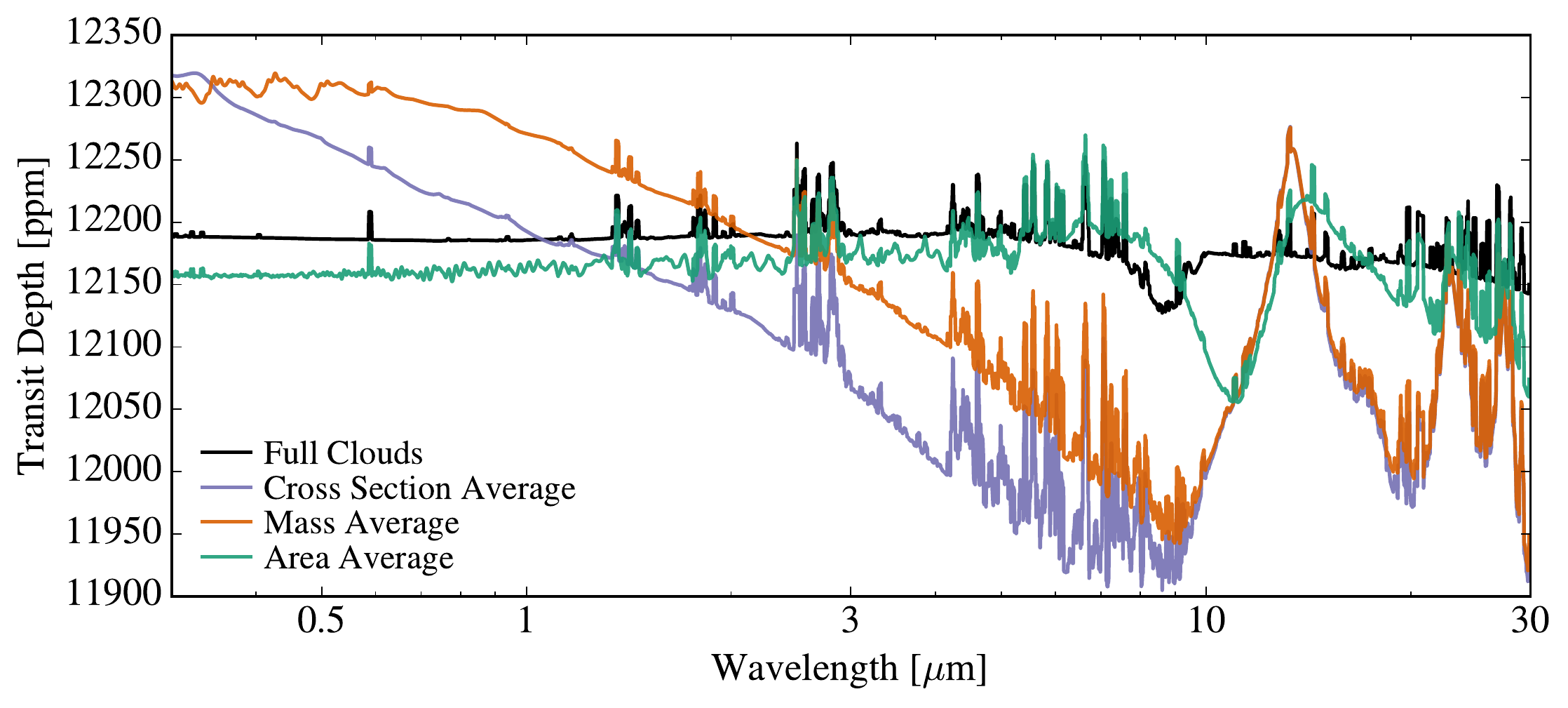}
\plottwo{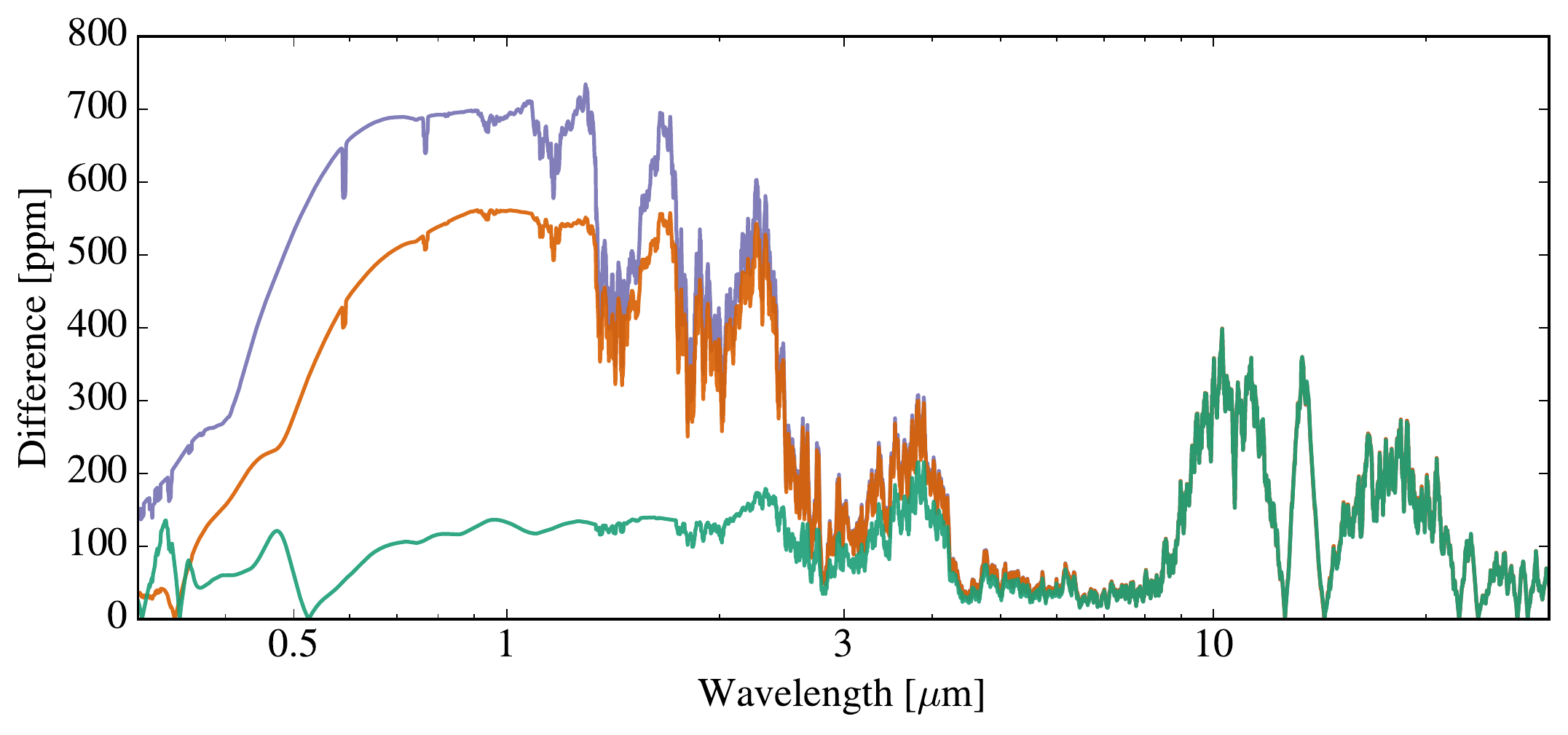}{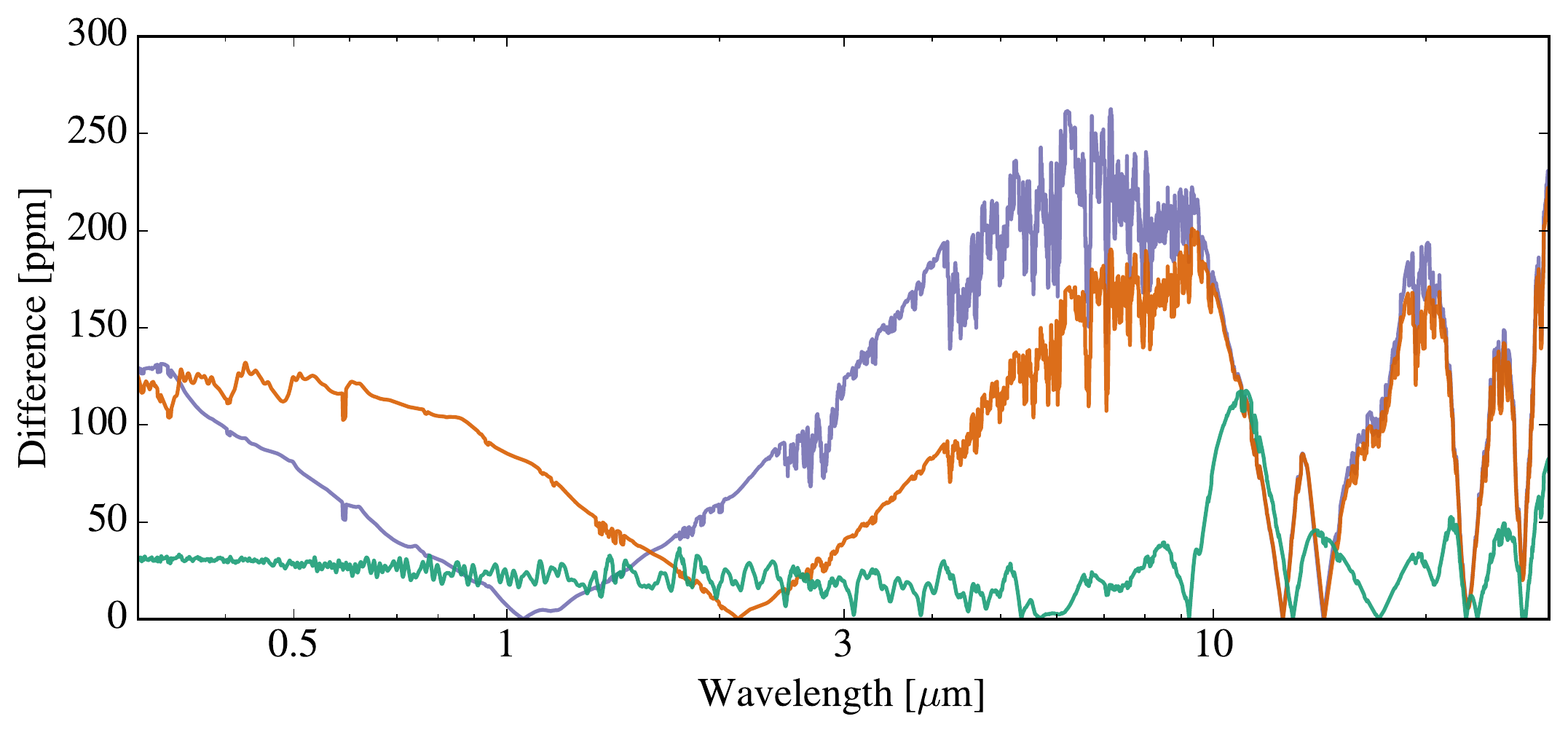}
\caption{The absolute value of the difference between considering the fully resolved cloud particle size distribution (black spectrum) and assuming a mean particle size with the same cloud mass (blue, red, and green spectra) can be as large as 700 ppm. Here we show the 2100 K planet at the east (left) and west (right) limbs andthe difference between the full size distribution and a calculated mean size (left, bottom).}
\label{meansize}
\end{figure*}

\subsection{Sensitivity to Cloud Particle Size Distributions}

Considering the fully resolved cloud particle size distribution is essential when calculating observed transmission spectra and interpreting observations. To demonstrate this importance we compare the derived transmission spectra using calculated \textsc{CARMA} cloud opacities to a transmission spectra calculated using a single representative cloud particle size for each species at each atmospheric height. In particular, we consider the case in which the total condensed cloud mass is conserved. We calculate the average particle size through averaging the full \textsc{CARMA} cloud particle size distribution weighted by particle cross-section ($\pi r^2$), area ($\pi r^3$), or mass. 

Transmission spectra calculated using a fully resolved particle size distribution differ distinctly from those calculated using a mean particle size for planets in our sample. Two examples of this effect are shown in Figure \ref{meansize}. While the area-averaged particle size is the closest to matching the opacity of the full particle size distribution, we find large differences in the transmission spectra across the entire broad wavelength range for every mean particle size probed. The largest difference in calculated transmission spectra can be as much as 700 ppm and is typically on the order of several hundred ppm. This difference is quite typical for the planets probed in this sample \footnote{In these cases, the difference between the limbs remains large and can be as much as 1000 ppm across a significant wavelength range (Figure \ref{meansize}).}. In particular, the change in the optical slope and relative strength of the infrared cloud features demonstrate that these regions of the spectra are particularly sensitive to the distribution of cloud particles. For example, using a mean particle size gives rise to a sharp silicate feature, which is broadened when considering a distribution of particle sizes. 

 Using a representative particle size instead of a full particle size distribution will lead to an incorrect interpretation of cloud properties. Furthermore, reducing the cloud particle size distribution to a single representative size will likely skew retrieved planetary properties and abundances as a single representative particle size is not able to reproduce the spectra over a broad wavelength range, particularly the broad cloud features in the infrared. The difference in transmission spectra will also be significantly larger if other methods of calculating cloud properties do not estimate the correct total cloud mass, species, or the location of the cloud particles in the atmosphere. It is therefore essential to accurately model cloud properties when interpreting observations to characterize planetary atmospheres.

\begin{figure}[tbp]
\epsscale{1.15}
\plotone{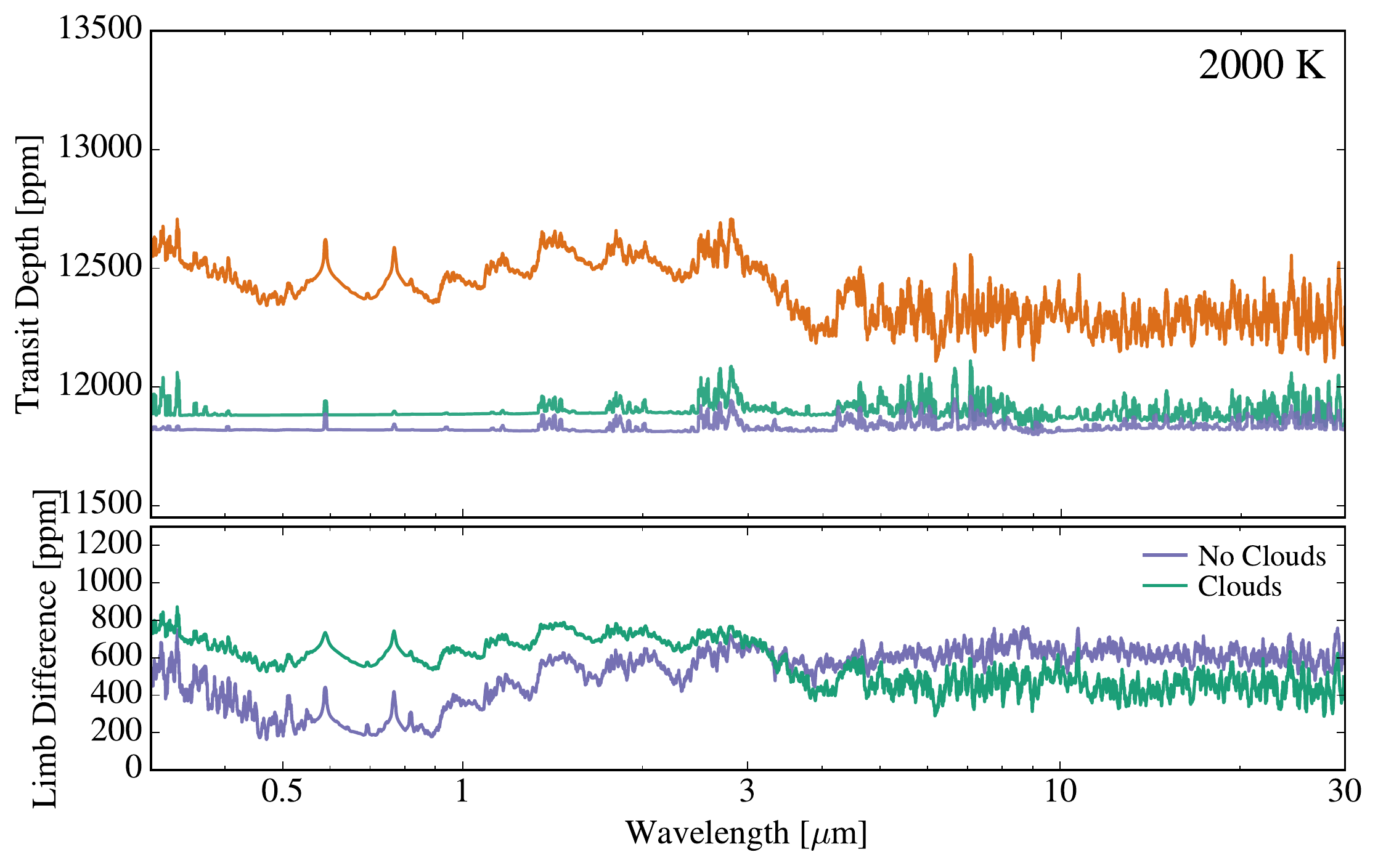}
\caption{Same as Figure \ref{max_trans} for microphysical parameters that lead to less cloud formation for a hot Jupiter with T$_\text{eq}$ = 2000 K. The spectra at the east limb appears significantly less cloudy than the spectra for the same object with different microphysical parameters shown in Figure \ref{max_trans}.}
\label{min_trans}
\end{figure}

\subsection{Sensitivity to Microphysical Parameters}\label{micro_sensitivity}

To demonstrate the sensitivity of our results to the species' desorption energy and contact angle, we simulate cases where all modeled species have roughly the minimum desorption energy of 0.1 eV which is representative of the desorption energy of small molecules (such as CH$_4$) from silicate grains \citep{1983Ap&SS..94..177S,2015MNRAS.454.3317S,2017MNRAS.472..389S}. For the contact angle, we approximate the work of adhesion following the geometric mean method \citep{owens_wendt} assuming that the surface energy of a given species is made up entirely of either dispersive or polar contributions. The contact angle is therefore calculated as: $\cos{\theta_c} = W_{C,x}/\sigma_x-1$ where $W_{C,x}=2\sqrt{\sigma_x\sigma_C}$. This method of estimating the contact angle provides a value that is smaller than the true contact angle, if the surface tension of the cloud condensation nuclei and/or the condensing species is made up of both polar and dispersive contributions as is common for most species, though larger than the angle used in our nominal setup. These changes to the desorption energy and contact angle result in less efficient cloud formation, particularly for species that form on cloud condensation nuclei. 

We find that our results are sensitive to these microphysical parameters, primarily because the efficiency of silicate and aluminum cloud formation is reduced. This effect can be most readily seen at the marginally supersaturated east limbs of the model atmospheres for the hotter planets in our grid where the molecular features are significantly less damped by clouds as shown in Figure \ref{min_trans}. Furthermore, in this setup, chromium and iron clouds no longer form on the west limbs. As these species do not impact the resultant transmission spectra, however, this change does not result in spectra on the west limbs and poles that are significantly different from the nominal cases shown in Figure \ref{max_trans}.   

To increase the accuracy of predictions from cloud microphysics in the future, the exact value of a species' contact angle and desorption energy needs to be determined from laboratory experiments. 

\section{Synthetic Light Curves and Observability of Light Curve Signatures}\label{observability}

With the next generation of instruments on the horizon it is of interest to know whether, and with what certainty, the presence of inhomogenous clouds could be detected directly on exoplanets through the transit method. 
Not only will the final spectrum be imprinted with the signature of clouds, but a time variable signal will be present in the transit lightcurve, as different regions of the planet's atmosphere are preferentially weighted throughout the course of the transit, particularly during ingress and egress, when only one terminator of the atmosphere is transiting. This time varying transit signal has been used to detect the presence of high velocity equatorial jets on exoplanets \citep{louden2015spatially}, and would also be sensitive to inhomogenous cloud coverage.

\begin{figure}[tbp]
\includegraphics{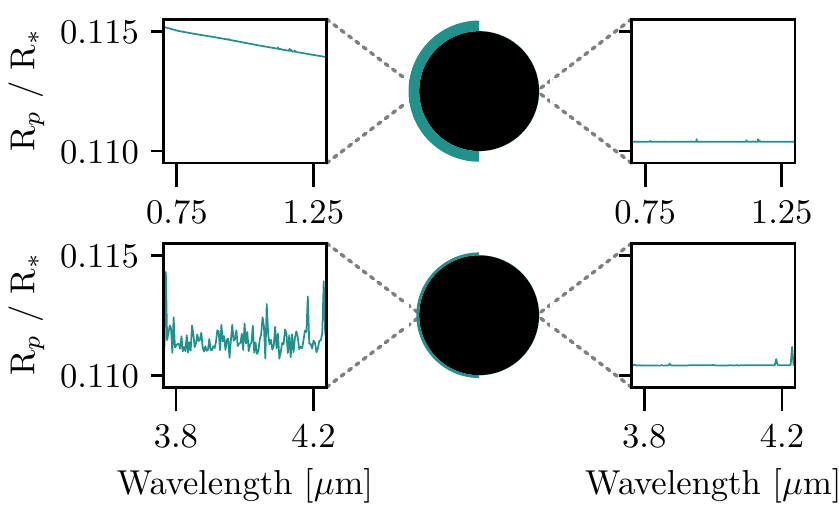}
\caption{A comparison of the spectra on the eastern and western sides of 2100 K planet with clouds at short (\emph{Top}) and long (\emph{Bottom}) wavelengths, and a scale diagram showing the resulting difference in radius (highlighted in green) which forms the input of the \textsc{terminator} model. The scale of the atmosphere has been increased by a factor of 5 for clarity. In this case, the asymmetry of the planet is significantly greater at short wavelengths due to the scattering slope feature.}
\label{terminator1}
\end{figure}

We simulated the time variable transit signal using the code \textsc{terminator} \citep{Louden2019}, which uses the same framework as \textsc{spiderman} \citep{Louden2018}, but modified for use on transits rather than secondary eclipses. Both codes use a geometric algorithm for calculating analytically the area obscured by the occulting object at each point in time during a transit/occultation. For this work, the algorithm was used to simulate the shape of a lightcurve when the opacity of the atmosphere varies around the limb. The planet is represented as a circle, with an additional half-annulus of variable width to represent the additional `height' of the atmosphere, either due to absorbing species blocking light in lower pressure regions of the atmosphere, or locally higher temperatures increasing the scale height and leading to a locally more extended atmosphere. A schematic diagram of the model is shown in Figure \ref{terminator1}.

We first describe the simulated observations and the observational consequences of inhomogenous clouds through a simple forward model. We will then go on to show that the inference of inhomogenous clouds from these simulated observations is statistically robust in retrieval, even with limited wavelength coverage and in the presence of uncertainty on limb darkening coefficients and imprecise transit times.

\subsection{Forward model}

\begin{figure*}
    \centering
    \epsscale{0.55}
    \plotone{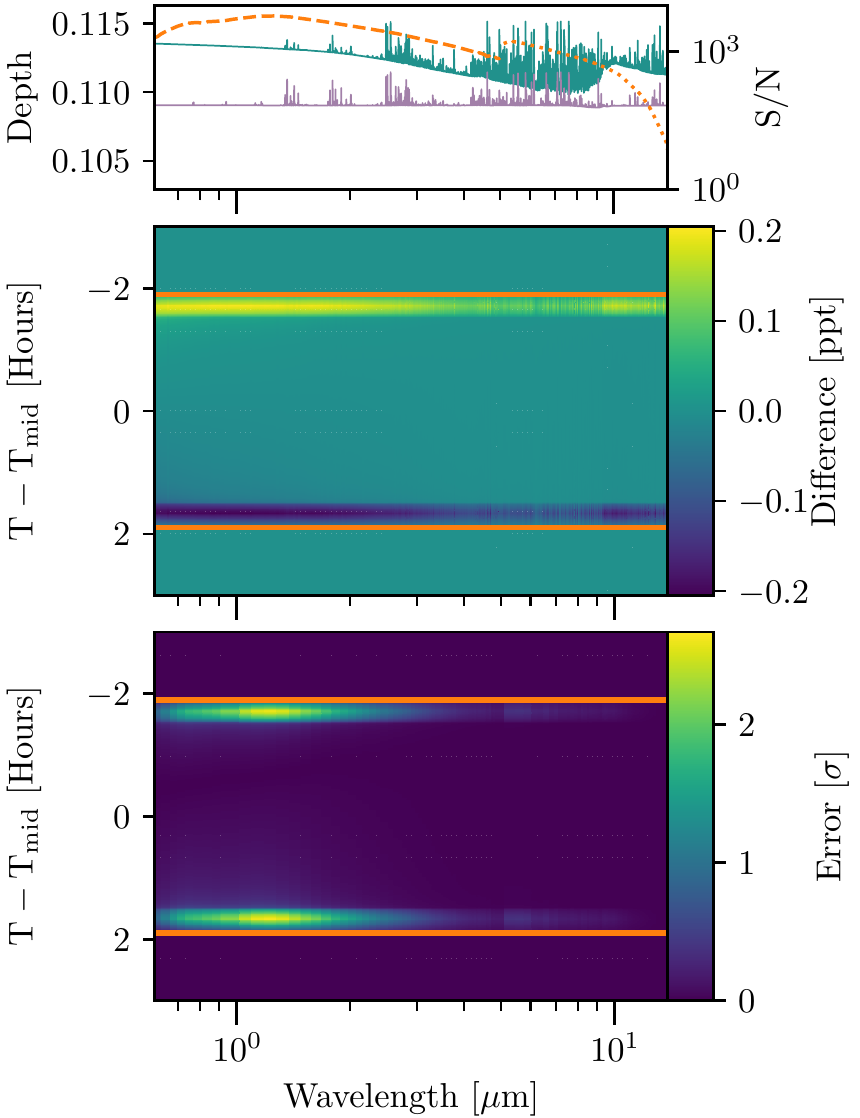}
    \plotone{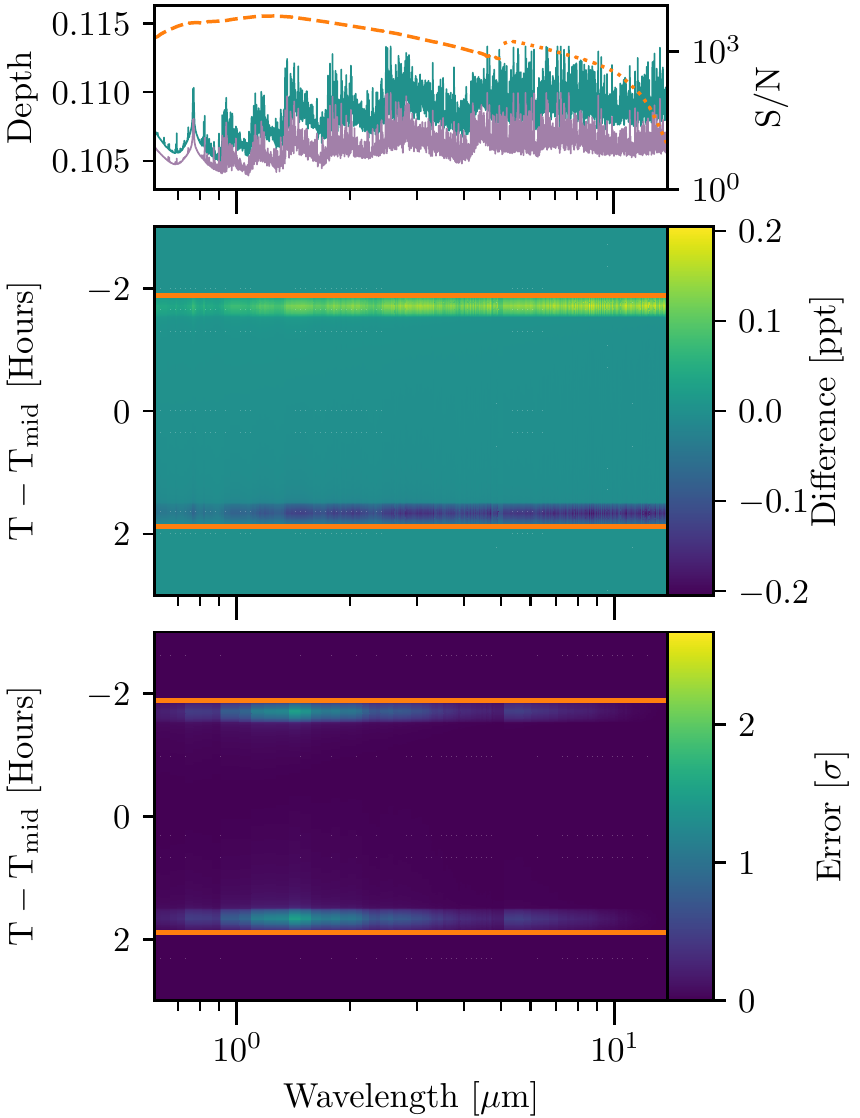}
    \caption{Simulated observations of asymmetric planets for a 2000 K planet, with a cloudy case on the left and a model with no clouds on the right. \emph{Top}: The transit depths for the eastern (green) and western (purple) terminators of the planet, plotted alongside the S/N of the simulated JWST observation in orange, dashed for NIRCAM and dotted for MIRI LRS. \emph{Middle}: The absolute difference between assuming a planet with an asymmetric atmosphere and a uniform one as a function of time and wavelength. \emph{Bottom}: Detectability of the signal with JWST for a planet around a star with the same magnitude as HD\,209458. The wavelength channels have been binned by a factor of 5 for clarity.}
    \label{forward}
\end{figure*}

We simulated a hypothetical observation of a hot Jupiter with \textit{JWST} using PandExo \citep{batalha-2017}, using NIRCAM and MIRI LRS to cover the wavelength range from 0.6 to 12 micron. Short exposure times of 10 seconds are used to capture the highest amount of information on the shape of the transit. As a test scenario, we assume a star-planet system similar to HD\,209458 (G type star, J mag 6.6) with a planet orbiting with an inclination of 90 degrees for a realistic `best case' observable target. The simulated observations with PandExo are used to calculate the signal to noise on each exposure in the lightcurves generated by \textsc{terminator}.

\textsc{terminator} requires a stellar model to implicitly account for the limb darkening at every wavelength. A limb-resolved model of the star was calculated using Spectroscopy Made Easy \citep{sme} with 99 limb angles sampled.

For each of the simulations described in section \ref{sim_cases} we first calculated a transit model for a planet with a uniform atmosphere, which is constructed by averaging the east and west terminator models. The uniform model is compared to a lightcurve made using inhomogenous atmospheres. In all cases there is a very clear difference between the two resulting lightcurve models, which can easily be seen by subtracting one from the other, as shown in Figure \ref{forward}. As expected, the difference is largest at ingress and egress, where one terminator is much more heavily weighted than the other in the inhomogenous case. With the simulated signal to noise for a JWST observation the difference between the two models is statistically highly significant, with over $2 \sigma$ of difference in some individual wavelength channels.

\subsection{Retrievals}

\begin{figure}[tbp]
\includegraphics{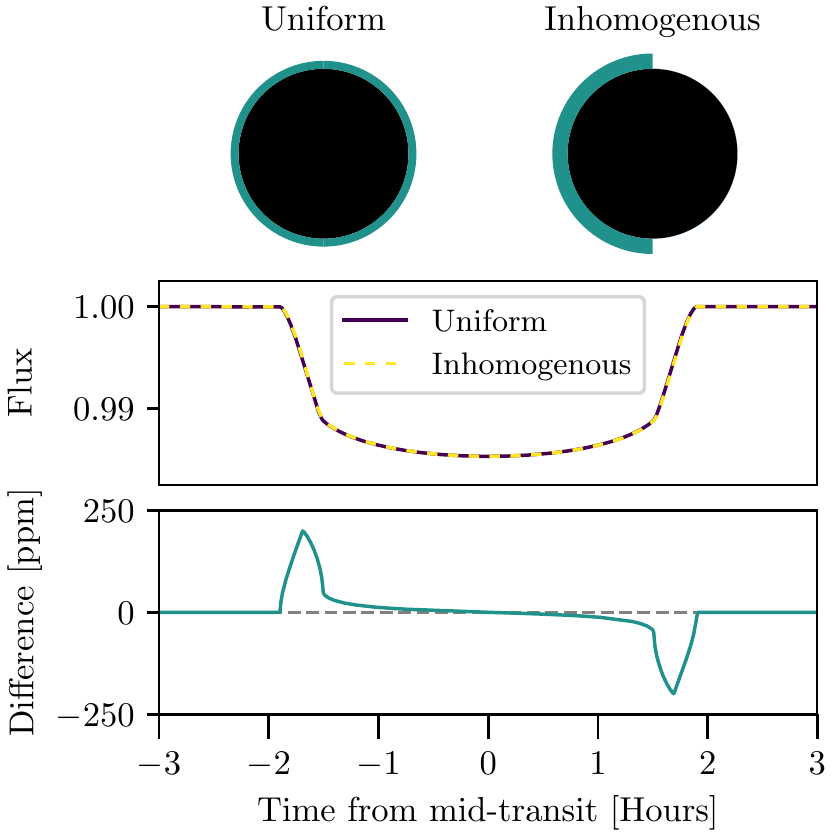}
\caption{\emph{Top:} Diagram of the planet models used for the 2100 K case, where the additional atmosphere height is highlighted in green and has been inflated by a factor of 5 for clarity. \emph{Middle:} The lightcurves calculated by TERMINATOR for these planet geometries \emph{Bottom:} The difference between the two lightcurves - the presence of an asymmetric atmosphere leads to a characteristic signature, similar to the residuals from an incorrect ephemeris.}
\label{terminator2}
\end{figure}

Figure \ref{terminator2} shows the difference in the shape of the lightcurve between the homogeneous and inhomogeneous models. The results are similar to those of \citet{von-paris-etal-2016}, who showed that the observational consequence of a planet with a different absorption radius on the eastern and western limbs is a distorted lightcurve, which to first order looks very similar to what one would expect if the ephemeris of the planet were not known accurately enough, presenting a slightly early or late transit. This time offset would typically be small, on the order of a few tens of seconds, so it would seem difficult to confidently assign this to the atmosphere of the planet instead of an error in the calculated ephemeris of the planet. However, as can be clearly seen in Figure \ref{forward} the effect is chromatic, in that it is largest at wavelengths where there is a large difference in the spectra, and smallest where the two sides are more similar.

It is therefore possible to confidently distinguish between uncertainty on the ephemeris and a true asymmetric atmosphere signal if a chromatically resolved signal is found where larger offsets correspond in wavelength to expected features in the planetary spectrum. This conclusion would be strengthened further if combined with optical phase curve observations of the planet that indicate inhomogenous cloud coverage, such as a westward offset bright spot in the optical \citep{Dang-2018}, or a transmission spectra showing clear signatures of clouds in the atmosphere (see Section \ref{trans_features}).

We ran a full Bayesian recovery test to check whether it was possible to infer differences in the two sides of the planet from a low resolution spectral lightcurve even in the presence of uncertainty on the true ephemeris of the planet, and to test whether this could be attributed to inhomogenous cloud cover. While an ideal retrieval would use the entire wavelength range, we use a minimal case of two small wavelength bands for clarity, in order to isolate the observational signal and potential confounding variables, therefore the significances we find in this section should be considered lower limits on what are possible.

The planet is observed for a full transit in the two wavelength regions, $\lambda1$ and $\lambda2$ - First with an instrumental setup optimized to observe the wavelength region with the largest expected distortion in the lightcurve, and a second instrumental setup optimized for the wavelength of the smallest effect to establish a baseline.

A simple metric is used to select $\lambda1$, $\lambda2$ and the optimum JWST instruments used in the retrieval. Using the model transit depths the difference between the eastern and western limbs is calculated for both the cloudy and the cloud-free cases, $d1$ and $d2$. Since the purpose of this test is to discriminate between these two cases, we then take the difference between $d1$ and $d2$, and multiply it by the signal to noise of each JWST instrument available on pandexo, therefore the metric, $m$ for a JWST instrument i is  
\begin{equation}
m_i = (d1-d2) \cdot SNR_i
\end{equation}
$m$ is then optimised over both instrument and wavelength.
The second region, the baseline, is chosen similarly with metric $m2$. If $d1_{max}$ is the value of $d1$ where $m$ is optimized, then
\begin{equation}
m2_i = (d1-d1_{max}) \cdot SNR_i
\end{equation}
In all cases tested the greatest signal to noise was achieved with $\lambda1$ having a short wavelength (~1 micron) and being observed with NIRCAM, and $\lambda2$ having a longer wavelength (~6 micron) being observed with MIRI LRS. It is important to note that it is not a necessity to observe with two instruments, and the same results can be achieved with a single instrument, though with slightly lower significance.

When forward modelling we used a full stellar atmosphere model to account for limb darkening at all wavelengths. We wish to check that the uncertainty on the limb darkening parameters does not correlate with any measures of inhomogeneity during retrieval, and since it is computationally expensive to calculate atmosphere models we instead use a quadratic limb darkening law. The limb darkening coefficients for the two wavelength regions are calculated using ldtk \citep{Parviainen-2015}. ldtk provides an uncertainty on the limb darkening parameters propagated from the uncertainty of the parameters on the star, which we conservatively increase by a factor of 10 and use as priors in our retrieval. We use a modified version of ldtk to work in the triangular sampling re-parametisation of \citet{kipping2013} for increased efficiency.

 \begin{figure*}[tbp]
\plotone{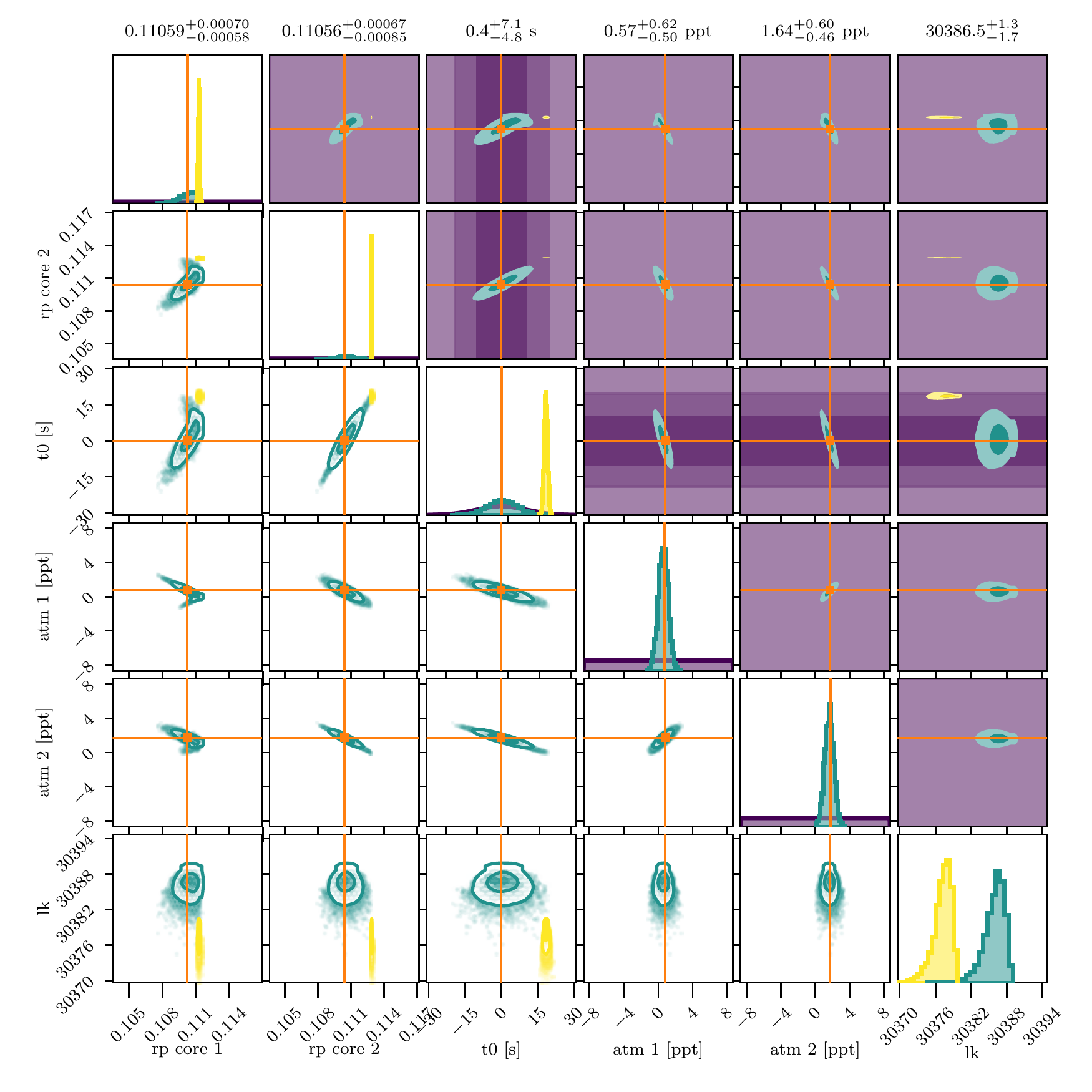}
\caption{An example corner plot from Bayesian retrieval for the 2100 K models. Purple indicates prior distributions used, the model posteriors are in green and the posteriors for the null model are in yellow. The values used to generate the model are indicated with orange lines. ``rp core'' 1 and 2 are the radii of the planet in the two wavelength bands before additional segments are added, ``t0'' is the difference in time of central transit from the prior value, ``lk'' is the log likelihood of the model. The addition of the parameters atm 1 and 2, the asymmetric atmosphere area for the two wavelength regions, significantly increases the quality of the fit.}
\label{corner}
\end{figure*}

 \begin{figure*}[tbp]
\includegraphics{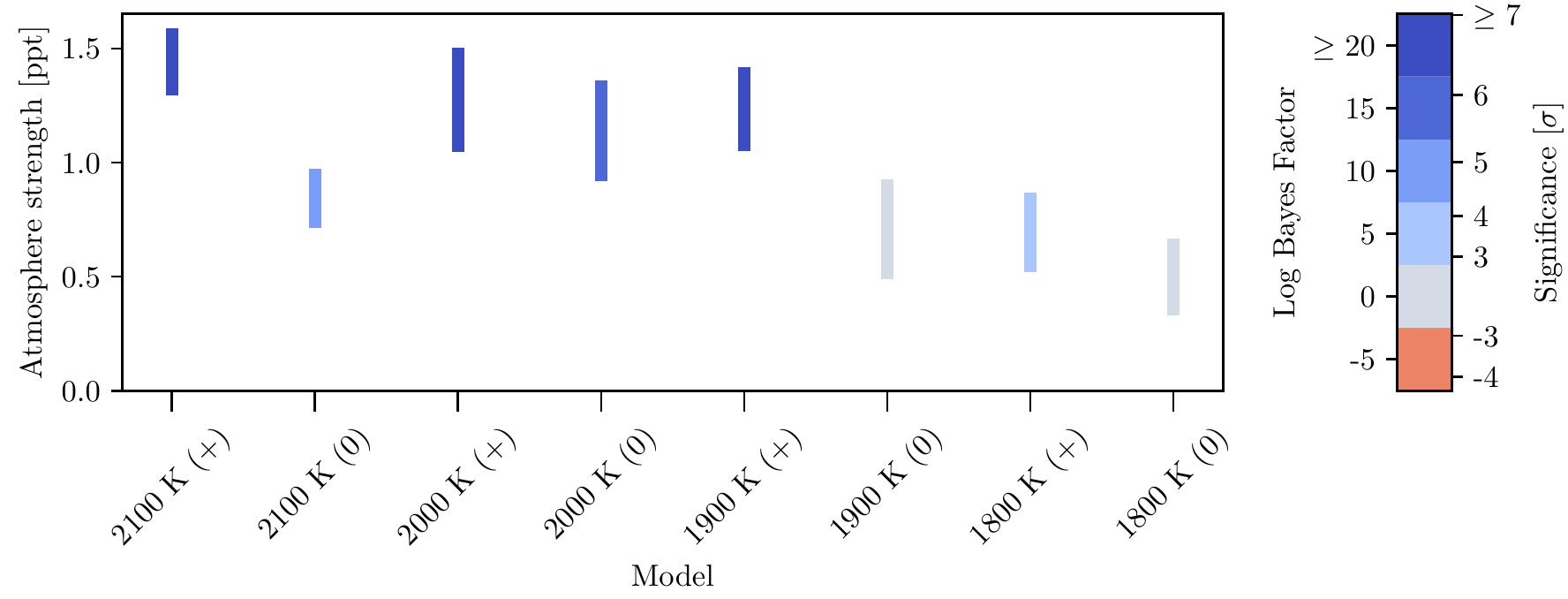}
\caption{The detectability of the lightcurve asymmetry in the test system as a function of equilibrium temperature. Cases marked with a (+) %or (-) 
are those where clouds are present at their maximal %and minimal predicted 
level%s
, those marked with a (0) are where no clouds are calculated. Atmosphere strength is the modulus of the ``atm'' value for the wavelength with the strongest asymmetry effect, i.e., it is the additional fractional area of the star covered compared to a model with no additional atmosphere. The length of the bar is the 68\% credible interval. The color code is a sigma-equivalent of the Bayes factor for how favoured the asymmetrical model is in each instance to the uniform one. Blue indicates strong evidence for the more complex model, grey that the Bayes Factor was too small to make a strong inference, and red that the simpler model is preferred.}
\label{retrieval_sig}
\end{figure*}

For each planet case we generate a simulated inhomogenous lightcurve for the two wavelength regions with the noise from pandexo, and then attempt to fit these lightcurves and recover the parameters used to initially generate the model. As part of the Bayesian analysis, we also fit a null hypothesis model, where the atmosphere of the planet is uniform, as would typically be assumed. Comparing the Bayesian evidences of these two scenarios gives the Bayes factor, which determines whether the more complex two-sided planet model is justified by the data. The retrieval is calculated through nested sampling with PyMultinest \citep{Buchner-2016}.

In our retrieval we assume the planet is spherical (i.e. we do not consider oblateness) and fix the planet's orbital parameters with the exception of the time of central transit t0, which we assume is known with an a-priori precision of $\pm$ 10 seconds. Of the remaining fit parameters, the radius of the planet before any additional atmosphere segments are added is denoted ``r$_p$ core'' 1 and 2 for $\lambda1$ and $\lambda2$. The additional absorbing area due to the atmosphere segment ``atm'' 1 and 2 in units of fractional stellar area, and is positive to denote it is on the eastern limb and negative to denote the western limb. A completely uniform atmosphere would have a value of zero for atm 1 and 2. The planet radius and atmosphere area for both wavelengths have uniform priors.

This model is compared to a `null' model, which has identical priors but lacks the parameters for the additional atmosphere segments on top of the planet base, i.e., it is a standard transit model. The asymmetric atmosphere model has a total of nine fitted parameters, and the null therefore has seven.

The advantage of using Multinest over other methods such as MCMC is that it calculates the Bayesian evidence, allowing rigorous model selection rather than relying on approximations such as the Bayesian Information Criterion (BIC). The comparison of the Bayesian Evidences of the two scenarios gives the Bayes Factor, which gives an assessment of which model is best supported by the data, weighted by the model complexity. 

Inspecting an example corner plot, Figure \ref{corner}, for the case of a 2100 K atmosphere with clouds shows that as expected, the asymmetric atmosphere parameters correlate with the time of central transit for both wavelengths. However, since the central transit time is shared by the two wavelength regions the degeneracy is lifted - i.e. attempting to fit an observation of an asymmetric atmosphere with a model with zero asymmetry in the atmosphere would require a different value of t0 in the two wavelength regions, since this value must be shared the fit is poor. This same breaking of the degeneracy worked in every case tested. 

The results are shown in Figure \ref{retrieval_sig}, with both the strength of the detected atmosphere and the significance over the null model shown. We found that in all of our tested cases the preferred excess atmosphere depth was significantly greater than 0, and the Bayesian evidence with respect to the null hypothesis of a uniform atmosphere was greater than 3 $\sigma$ equivalent in all but 2 of the tested scenarios, both of which were cloud free model atmospheres, at 1800 and 1900 K. %maybe "both of which were cloud-free model atmospheres?" Also, I'm loving these Sections!

The evidence for an inhomogeneous atmosphere and strength of the feature was in all cases significantly stronger when clouds were included in the atmosphere model, and increased in strength with the planetary equilibrium temperature. This technique is therefore capable of robustly detecting inhomogeneity in a planetary atmosphere with JWST, even in the presence of uncertainty on the time of central transit and limb darkening parameters. Inhomogenous cloud cover could be studied in detail by combining this technique with optical phase curves, providing strong constraints on cloud formation models.

\section{Discussion}\label{discuss}

\subsection{Implications for Interpretation of Phase Curve Offsets}

The signatures of inhomogeneous cloud cover discussed in Section \ref{observability} will help in both understanding limb averaged observations of inherently three dimensional planetary atmospheres and constraining interpretations of exoplanet phase curves.

Exoplanet phase curves are particularly powerful tools that can give insight into inhomogenous cloud cover, chemistry, and temperature structure, particularly when observed at multiple wavelengths \citep[e.g.,][]{2018haex.bookE.116P}. However, these observations are intrinsically difficult to make as both the timescale ($\sim$ days instead of hours) and shape of a phase curve makes these observations more difficult than occultations \citep{2017PASP..129g2001S,2018haex.bookE.116P}. In addition to the intrinsic observational difficulties of observing phase curves, there are significant interpretation difficulties such as the higher uncertainties in derived atmospheric properties \citep{2015AJ....150..112S}.

The combination of the transit signatures of inhomogenous clouds presented in this work will aid in breaking the degeneracies between non-atmospheric and atmospheric contributions to the planetary phase curve as the presence of inhomogeneous clouds can be constrained by this complementary and cheaper method.  

\subsection{Tests for Mechanisms that Could Reduce Cloud Inhomogeneity}
While inhomogeneous cloud cover on the limbs of hot Jupiters with T$_\text{eq} \sim 1800 - 2100$ K is a likely outcome, the models described in Section \ref{cloud_model} and the observational metrics described in Section \ref{observability} will allow us to test for the following three mechanisms that could reduce inhomogeneities in limb cloudiness. 

Firstly, it is possible that atmospheric circulation at the equilibrium temperatures probed in this study is strongly affected by atmospheric drag. In this case, the large scale atmospheric winds flow from day to night \citep{showman-etal-2013}. Such a flow pattern may equilibriate the temperature structure at the limbs and lead to a homogenized cloud population. Thus, the metric presented in Section \ref{observability} could indirectly probe the flow pattern, and thus the strength of the atmospheric drag, if there is strong evidence for a homogeneous model for a hot Jupiter in this temperature regime. 

Secondly, the metric presented in this work could test for the presence of a condensible species in the upper atmosphere of the west limbs, such as MnS, which may mimic the high-altitude silicate and aluminum clouds present on the east limbs at short wavelengths. Although MnS clouds are not thought to form abundantly \citep[Section \ref{dominant_clouds},][]{Gao2019}, this theory could be tested observationally using such a probe of inhomogeneity. This mechanism could be distinguished from the first mechanism with observations at longer wavelengths where the east limb appears more clear or if broad features due to MnS or other cloud species are present in the transmission spectrum. 

Finally, this metric could test for the presence of homogeneous high level hazes that can be transported across the planet and can form and persist at high temperatures. Though the formation pathway of hazes is an active area of research, this mechanism may be distinguishable from the first two mechanisms because the production of photochemical hazes may depend on the energetic flux from the host star and should thus vary for different planets as a function of this quantity. Indeed, many hot Jupiters do not show significant evidence of high level hazes that could produce a significant Rayleigh-like slope in the optical \citep[e.g., ][]{2016Natur.529...59S} suggesting that at least some hot Jupiters are not homogeneously covered in high temperature hazes. 

\section{Conclusions}\label{conclude}
We simulated clouds on hot Jupiters with 1800 K $<$ T$_\text{eq}$ $<$ 2100 K using a size distribution-resolving cloud microphysics model to assess the feasibility of observing inhomogenous clouds from transit observations alone. Cloud formation is efficient for all planets probed in the modeled grid. The model transmission spectra including microphysical clouds is different on each limb of the planet, often by as much as $\sim$1000 ppm. At short wavelengths, despite having lower total cloud mass, the east limb appears cloudier than the west limb for planets with equilibrium temperatures less than 2100 K. Silicate clouds typically dominate the cloud opacity for all planets in our model grid with the exception of the hottest planets at the east limbs where aluminum clouds also significantly contribute to the total cloud opacity. 

There are three primary transmission spectrum signatures of condensational clouds. First, condensational clouds can substantially mute absorption features across a broad wavelength range. Second, clouds can also contribute to muted and sloped spectra in the optical. The strength of this slope depends strongly on temperature with hotter planetary locations producing more distinct slopes. Third, both silicate and aluminum clouds can also give rise to broad features \citep{2015A&A...573A.122W} in the infrared between $\sim$ 10 - 20 $\mu$m that should be observable with \textit{JWST} \citep{Venot2019,2017ApJ...850..121M}. Furthermore, for some locations in a planetary atmosphere we find a cloud-free spectral window from $\sim$ 5 - 9 $\mu$m where the opacity of silicate clouds is decreased. 

The cloud paerticle size distributions in this work are not log-normal and differ for different condensible species. It is essential to use the full cloud particle size distribution when interpreting or creating model spectra, as considering a representative particle size leads to a spectra that is unpredictably and significantly different in both magnitude and shape - though even these cases produce spectra that are different between the east and west limbs. Furthermore, the cloud content of a given planet is sensitive to the material properties of a condensible species - namely its desorption energy and contact angle. 

We use the fact that the observed difference in limb radii in the presence of clouds characteristically changes with observing wavelength to assess the feasibility of observing inhomogeneous clouds in transmission with \textit{JWST}. We use the detailed \textsc{terminator} code to map the transits of the modeled planets including inhomogeneous limb radii. Using a forward model across a broad wavelength range, the errors in fitting a homogeneous model to observations of a planet with inhomogeneous clouds leads to chromatic errors that are distinct from the clear case - thus providing a clear signature of inhomogeneous clouds. Using an inverse bayseian retrieval, we show that these synthetic JWST observations can be used to probe inhomogenous clouds in a way that is statistically robust, even with limited wavelength coverage and in the presence of uncertainty on limb darkening coefficients and imprecise transit times.

\section{Acknowledgements}
We would like to thank N\'estor Espinoza, Natasha Batalha, and Hannah Wakeford for their insightful discussion. This material is based upon work supported by the National Science Foundation Graduate Research Fellowship under Grant DGE1339067. X.Z. acknowledges support from NSF grant AST1740921. P.G. acknowledges support from the 51 Pegasi b Fellowship funded by the Heising-Simons Foundation. T.L. gratefully acknowledges support from the David and Claudia Harding Foundation in the form of a Winton Exoplanet Fellowship.

\bibliography{refs}
\bibliographystyle{aasjournal}

\appendix
\section{Condensible Species}\label{species}
As an update to the model presented in \citet{2018ApJ...860...18P}, we consider a suite of six different condensible species: TiO$_2$, Al$_2$O$_3$, Fe, Mg$_2$SiO$_4$, Cr, and MnS. Three of these species, TiO$_2$, Fe, and Cr, are able to form through either homogeneous nucleation or heterogeneous nucleation due to the presence of these molecules in the gas phase. TiO$_2$ clouds homogeneously nucleate abundantly due to their very low surface energies, and TiO$_2$ can thus serve as cloud condensation nuclei for other cloud species \citep[see][for more details on cloud condensation nuclei at high temperatures]{2018A&A...614A.126L}. Since homogeneous nucleation is very efficient for this species, heterogeneous nucleation is not their favored pathway for formation. Thus, only Fe and Cr cloud particles can form via homogeneous or heterogeneous nucleation pathways, though heterogeneous nucleation is typically favored if an abundance of cloud condensation nuclei are present due to their relatively high surface energies. Al$_2$O$_3$, Mg$_2$SiO$_4$, and MnS likely form via grain chemistry as they do not exist in this form as gases. We therefore model their formation as a parameterization of heterogeneous nucleation theory following \citet{2006A&A...455..325H} and detailed in \citet{Gao2019}. For all cloud species considered in this work we assume an initial solar concentration of gaseous species in an abundance given by equilibrium chemical modeling before rainout for that planetary temperature profile as described in \citet{Gao2019}.

These cloud species will form when they reach a sufficient supersaturation in the atmosphere. This is defined as the point where a species' atmospheric partial pressure exceeds its saturation vapor pressure. This can be converted to a species' condensation curve as detailed in \citet{2012ApJ...756..172M} where we expect cloud particle growth to occur most rapidly at locations where the condensation curve intersects a planetary pressure temperature profile \citep{2018ApJ...860...18P}. The condensation curves for the species considered in this work are shown in Figure \ref{pt_profs}. However, these curves are approximations for the cloud formation process and should only be used as a rough guide. This is because cloud formation will be highly inefficient for cases where supersaturations are not sufficiently large. This is demonstrated in \citet{2018ApJ...860...18P} for planets with equilibrium temperatures of 1800 K and 1900 K. From a naive consideration of the condensation curve, one might expect these planets will form titanium clouds at their substellar points. However, in these cases titanium clouds do not form significantly. This further highlights the importance of detailed microphysical cloud modeling.  

\section{Material Properties Used in Microphysical Modeling}\label{props}
There are several material properties needed when modeling clouds from first principles with microphysics. In particular these are: latent heat of vaporization, desorption energy, the species' surface energy, and contact angle (for heterogeneous nucleation). 

To derive a given species' latent heat, we use the Clausius-Clapyron relation in conjunction with the species' saturation vapor pressure \citep[see][]{Gao2019}. We use this estimated latent heat of vaporization estimate to derive a species desorption energy (see Table \ref{desorption}). 

A species's desorption energy regulates the rate of heterogeneous nucleation such that higher desorption energies result in more nucleation \citep[see][Appendix \ref{species}]{Gao2019}. This term is thus very important in regulating cloud formation processes. Based on experimental estimates for species not modeled in this work (such as H$_2$O), the desorption energy is typically on the same order, but less than, the latent heat of vaporization \citep{doi:10.1098/rsta.2002.1137,2013ctmc.book....3B,Hyunho2016}. In our modeling work we consider two different cases for the desorption energy as described in Section \ref{minmax}. 

The species' surface energies used in our modeling are determined either through detailed modeling or laboratory experiements. The specific values are given in \citet{Gao2019}. 

Finally, the contact angle used in modeling heterogeneous nucleation is determined through a consideration of the surface energy of the nucleating species and the cloud condensation nuclei (TiO$_2$ in this work). The contact angle regulates heterogeneous nucleation in a similar way to the desorption energy where larger contact angles lead to less nucleation. The maximum contact angle is 180$^o$ which results in no heterogeneous nucleation. The contact angle can be defined as:

\begin{equation}
    \cos{\theta_c} = \frac{\sigma_C-\sigma_{x,C}}{\sigma_x}
\end{equation}

\noindent where $\theta_C$ is the contact angle between the cloud condensation nuclei and the condensing species, $\sigma_C$ is the surface energy of the cloud condensation nuclei, $\sigma_x$ is the surface energy of the condensing species, and $\sigma_{x,c}$ is the interfacial tension between the cloud condensation nuclei and the condensing species. The interfacial tension is given by: $\sigma_{x,c} = \sigma_C + \sigma_x - W_{C,x}$ where $W_{C,x}$ is the work of adhesion which can be estimated by various methods. In our modeling, we consider two different cases for the contact angle of a species as described in Section \ref{minmax}.

 \begin{deluxetable}{lc}
\tablecolumns{2}
\tablecaption{Species Desorption Energies \label{desorption}}
\tablehead{   % column headings
  \colhead{Species} &
  \colhead{Approximate Desorption Energy} 
}
\startdata
TiO$_2$     & 3.22 eV \\  
Mg$_2$SiO$_4$ & 3.223 eV  \\
Al$_2$O$_3$ & 4.553 eV  \\
Fe & 2.083 eV \\
Cr & 2.034 eV \\
MnS    &  2.362 eV  \\  
Na$_2$S   &  1.378 eV  \\
\enddata 
\end{deluxetable} 

\end{document}